\documentclass[journal]{IEEEtran}

\usepackage{amssymb}
\usepackage[cmex10]{amsmath}
\usepackage{amsmath}
\usepackage{amsfonts}
\usepackage{amssymb}
\usepackage{amscd}
\usepackage{stmaryrd}
\usepackage{subfigure}
\usepackage{cite}
\usepackage{multirow}
\usepackage{flushend}
\usepackage{tikz}
\usepackage{pgfplots}
\usepgflibrary{shapes}
\usetikzlibrary{arrows,shapes,chains,matrix,positioning,scopes}

\newlength\tikzheight
\newlength\tikzwidth

\newtheorem{remark}{Remark}

\newcommand{\abs}[1]{\left\vert#1\right\vert}

\IEEEoverridecommandlockouts

\begin{document}

\title{Throughput Analysis of Primary and Secondary Networks in a Shared IEEE 802.11 System}
\author{Santhosh Kumar, Nirmal Shende,  Chandra R. Murthy, and Arun Ayyagari
  \thanks{Santhosh Kumar is with the Dept. of ECE, Texas A\&M Univ., USA. 
    He was with the Dept. of Electrical Communication Engg.~at the Indian Institute of Science (IISc), Bangalore, India during the course of this work. (e-mail: santhosh.kumar@tamu.edu)}
  \thanks{Nirmal Shende is with the Polytechnic Institute of NYU, USA. 
    He was with the Dept. of Electrical Communication Engg.~at the Indian Institute of Science (IISc), Bangalore, India during the course of this work. (e-mail:  nvs235@students.poly.edu)}
  \thanks{Chandra R. Murthy is with the Dept.~of Electrical Communication Engg.~at IISc, Bangalore, India. (e-mail: cmurthy@ece.iisc.ernet.in)}
  \thanks{Arun Ayyagari is with The Boeing Company, Seattle, USA. (e-mail: arun.ayyagari@boeing.com)}
  \thanks{This work was funded by a research grant from the Aerospace Network Research Consortium.}
}
\maketitle

\begin{abstract}  
  In this paper, we analyze the coexistence of a  primary and a secondary (cognitive) network when both networks use the IEEE 802.11 based distributed coordination function for medium access control. 
  Specifically, we consider the problem of channel capture by a secondary network that uses spectrum sensing to determine the availability of the channel, and its impact on the primary throughput. 
  We integrate the notion of transmission slots in Bianchi's Markov model with the physical time slots, to derive the transmission probability of the secondary network as a function of its scan duration. 
  This is used to obtain analytical expressions for the throughput achievable by the primary and secondary networks. 
  Our analysis considers both saturated and unsaturated networks.
  By performing a numerical search, the secondary network parameters are selected to maximize its throughput for a given level of protection of the primary network throughput. 
  The theoretical expressions are validated using extensive simulations carried out in the Network Simulator~2. 
  Our results provide critical insights into the performance and robustness of different schemes for medium access by the secondary network. 
  In particular, we find that the channel captures by the secondary network does not significantly impact the primary throughput, and that simply increasing the secondary contention window size is only marginally inferior to silent-period based methods in terms of its throughput performance.
\end{abstract}

\begin{IEEEkeywords}
  Cognitive radio, distributed coordination function, MAC protocols,  network throughput.
\end{IEEEkeywords}

\section{Introduction} 

The coexistence of a higher-priority \emph{primary} network with a lower-priority \emph{secondary} network is a well-studied topic in the literature (e.g., \cite{Zhu_2004, Ni_2004}).
The goal in these studies is to enable the secondary network to utilize the RF spectrum whenever possible, and at the same time, ensure that the negative impact on the performance of the primary network is minimal.  
A recent and emerging approach to dramatically improve the efficiency of spectral usage by a secondary network is that of Cognitive Radio (CR) \cite{mitola_cogradio_2000, haykin_brainempowered_2005, jafar_breaking_2009}. 
A CR device acquires knowledge of its radio environment through spectrum sensing, and tunes its transmission parameters to fully utilize the spectrum when it is idle, and cause minimal interference to the primary network when it is busy. 
Often, the CR proposals are targeted towards frequency bands (e.g., the Digital TV spectrum) where the primary signal does not employ any form of spectrum sensing prior to transmission.
However, the CR networks need to also be future-proofed against possible modifications to the physical and MAC layer protocols of the primary networks.
In particular, if the primary network employs a listen-before-talk MAC protocol such as the IEEE 802.11 distributed coordination function (DCF) \cite{gast_wirelessnets_2002, mohapatra2005ad, IEEE_std}, it is possible that the CR network could sense and find the spectrum idle even when the primary network is in its back-off or sensing phase.
Once the CR network starts transmission, the primary network will find the spectrum unavailable, and freeze its back-off timers.
After the CR network finishes its transmission, it senses the spectrum, and since the primary network is still in back-off, it finds the spectrum available again.
This undesirable phenomenon is referred to as \emph{channel capture} in the literature \cite{Liu_ISCCSP_2010}.
Typically, this is handled by either requiring the secondary network to exercise a mandatory silent period of duration exceeding the maximum possible back-off window of the primary network prior to any transmission, or by increasing the contention window length of the secondary network relative to that of the primary network.
However, these could result in a low throughput and inefficient use of the spectrum by the secondary network due to the high sensing duration overhead or due to the longer back-off overheads, especially when the primary network is lightly loaded.  

Thus, in the context of CR networks, one relatively unexplored area of research is the interplay between the network throughput of the primary/CR networks and the Medium Access Control (MAC) layer protocol, and is the focus of this work.
In particular, some pertinent questions to ask include: what is the effect of channel capture on the throughput of the primary network? 
How can the sensing and back-off parameters of the secondary network be tuned to maximize its throughput while offering a desired level of protection to the throughput of the primary network? 
How does the sensing-based spectrum access for CR compare with other MAC layer techniques such as simply increasing the secondary contention window length, in terms of CR throughput and primary network protection?

In this work, we are interested in analyzing the throughput performance of primary and secondary networks, when both networks employ the IEEE 802.11 DCF for medium access.
Our results are useful not only for understanding and predicting CR network performance under different scenarios, but also for designing the MAC layer parameters of the CR network to maximize its throughput while simultaneously satisfying an upper limit constraint on the loss of the primary network throughput. 
We also illustrate the generality of the analytical framework we develop by using it to derive the throughput performance of other simpler secondary MAC techniques such as increasing the contention window length and observing a mandatory silent period. 
For simplicity, we use the phrases secondary network and CR network interchangeably in this paper. 
We start with a survey of related literature.

References \cite{Zhu_2004} and \cite{Ni_2004} provide a survey of the QoS differentiation mechanisms in the context of IEEE 802.11 networks.  
In \cite{Liu_ISCCSP_2010}, the authors propose a sense-wait-transmit protocol for coexistence of CR networks with IEEE 802.11 WLANs. Here, the CR devices select their data transmit duration based on an estimate of the channel idle duration, to satisfy an upper bound constraint on the probability of collision with the primary network. 
In \cite{adamis_throughput_2009}, the performance of the intermittent DCF that allows for frequent spectrum scanning by the secondary network was analyzed, but that work does not account for the nonzero spectrum sensing duration of the CR network, or analyze the impact of the interweave access on the primary network.
A different approach was studied via simulations in \cite{Tanaka_CogArt_2009}, where the CR network randomly switches between an aggressive and a passive mode, such that the primary sees minimal change in channel conditions regardless of the secondary traffic conditions.
Also, in the aggressive mode, the secondary contends with the primary when it finds the channel to be idle, while in the passive mode, the secondary network senses the channel for a duration that exceeds the maximum primary back-off duration before transmission.
Dynamically adjusting the contention window length and other MAC layer parameters such as the number of back-off stages, inter-frame spacings, etc, based on the priority of the different users and the traffic conditions has been studied in \cite{Aad_Infocom_2001, He_LCN_2003, Banchs_WCNC_2002, Vaidya_2005}.
Other works on cognitive MAC protocols with a WLAN-type primary have appeared in \cite{Niyato_WirelessCommun_2009, shao-yi_hung_opportunistic_2008, qi_zhang_cognitive_2008, Lou_PIMRC_2009, sywang_enhancedmac_2011, Khabazian-2012}. With generic primary networks, MAC layer protocols have been proposed and their achievable throughput has been analyzed in~\cite{srinivasa_cognitive_2007, qzhao_decencogmac_2007, cordeiro_cogmac_2007, hangsu_crosscogmac_2008, Xu_WiCon_2008, Jiang_Infocom_2011, Park_TCOM_11}.

Thus, it is an interesting open problem to study the effect of the CR paradigm on a primary network that also incorporates a listen-before-talk protocol along with some form of DCF for medium access.
Such studies are important, because the future CR networks will be expected to perform well and have minimal deleterious impact on the primary, regardless of the MAC or higher layer protocols employed by the primary network. 
Our main contributions are:
\begin{enumerate}

\item We extend the Markov model in \cite{bianchi_performance_2000} to account for the intermittent spectrum sensing by the secondary network and characterize the effect of scan duration on the transmission and collision probabilities of the primary and secondary networks. (See Sec.~\ref{sec:saturated}.)

\item We provide the analysis for both the saturated and unsaturated networks. 
  In particular, for the unsaturated case, we propose a new Markov chain model that is consistent with \cite{bianchi_performance_2000} in the limit where the network is fully loaded. (See Sec.~\ref{sec:unsaturated}.)

\item We theoretically characterize the primary and secondary throughput performance under three different schemes for the secondary network: (a) periodic spectrum sensing and using a larger contention window than the primary network, (b) only using a larger contention window than the primary, and (c) observing a non-scanning silent period and a larger contention window than the primary. (See Sec.~\ref{pri_sec_thruput}.)

\item We use the theoretical development to numerically design the sensing duration and contention window size of the secondary network for a given set of primary network traffic conditions (packet arrival rates, number of nodes, etc) to maximize its throughput, subject to an upper limit constraint on the loss of the primary network throughput. (See Sec.~\ref{saturated_simulations}.)

\end{enumerate}
We use the popular network simulator (Ns-2) \cite{ns2} to show the excellent match between the theoretical expressions and the simulation results.
Through the experimental results, we find, somewhat surprisingly, that the channel capture effect does not significantly impact the primary throughput, provided the secondary scanning duration is at least of the order of a few idle slots.
Moreover, increasing the secondary back-off window is a simple and robust technique that offers nearly the same throughput performance as more complex methods.

\section{System Model} \label{sec:model}

\begin{figure}[t]
  \begin{center}
    \begin{tikzpicture}
  [scale=0.28,node distance=0.1mm,draw=black,thick, >=stealth',
  packet/.style={rectangle, minimum height=1pt, minimum width=3pt, draw=black,fill=gray}]  

  \def\TT{10};
  \def\t{3};
  \def\s{1.5};
  \def\T_2{5};
  \def\sh{2};
  \def\ph{1};
  \def\pl{2};

  \node at (-1.2,4) {\small{Primary}};  
  \node at (-1.2,3) {\small{Transmission}};
  \draw[fill=gray] (\T_2 - 2.9,\ph+3) rectangle (\T_2 - 2.9 +\pl,+3);
  \draw[fill=gray] (\T_2+\t+4.5,\ph+3) rectangle (\T_2+\t+4.5+\pl,3);
  \draw[fill=gray] (\T_2+\t+7.5,\ph+3) rectangle (\T_2+\t+7.5+\pl,3);
  \draw[fill=gray] (\T_2+\t+11.5,\ph+3) rectangle (\T_2+\t+11.5+\pl,3);
  \draw[fill=gray] (\T_2+\t+16.5,\ph+3) rectangle (\T_2+\t+16.5+\pl,3);

  \draw[thick,fill=red] (1.8,0.25) rectangle (\T_2+\t+\TT,0);
  \draw[thick,fill=green] (\T_2+\t,0.25) rectangle (\T_2+\t+\TT,0);
  \draw[thick,fill=red] (\T_2+\t+\TT,0.25) rectangle ((\T_2+\t+\TT+\t+\T_2,0);

  \node at (-1.2,0.25) {\small{Scanning}};
  \draw[thick] (1.8,0) -- (\T_2,0) -- (\T_2,\sh)  -- (\T_2+\t,\sh) -- (\T_2+\t,0) -- (\T_2+\t+\TT,0) -- (\T_2+\t+\TT,\sh) -- (\T_2+\t+\TT+\t,\sh) -- (\T_2+\t+\TT+\t,0) -- (\T_2+\t+\TT+\t+\T_2,0) ;

  \draw[thick,<->] (\T_2,\sh+0.3) -- node[midway,above=0.03]{$t$} (\T_2+\t,\sh+0.3);
  \draw[thick,<->] (\T_2+\t,0.9) -- node[midway,above=0.03]{$T-t$} (\T_2+\t+\TT,0.9);

  \node at (\T_2+\s,1) {$\mathit{idle}$};
  \node at (\T_2+\t+\TT+\s,1) {$\mathit{busy}$};

  \node at (-1.2,-2.5+0.5) {\small{Secondary}};
  \node at (-1.2,-2.5-0.5) {\small{Transmission}};
  \draw[fill=gray] (\T_2+\t+0.5,\ph-3) rectangle (\T_2+\t+0.5+\pl,-3);
  \draw[fill=gray] (\T_2+\t+4.5,\ph-3) rectangle (\T_2+\t+4.5+\pl,-3);
\end{tikzpicture}

    \caption{Timeline for the primary and secondary networks.
      The green regions indicate opportunities for secondary transmission.
      The red regions indicate that secondary network has frozen its transmission.
      Any overlap of the primary transmission during the secondary scanning results in a busy state.
    }
    \label{fig:timeline}
  \end{center}
\end{figure}
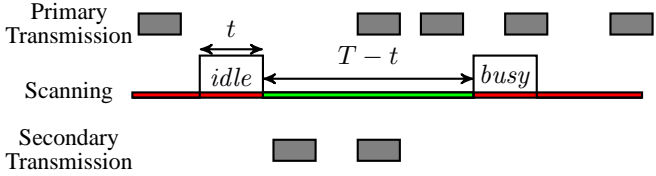

We consider a primary and a secondary network with $N_p$ and $N_s$ nodes, respectively. 
The primary network is an IEEE 802.11 based Wireless Local Area Network (WLAN), which uses the DCF to access the medium. 
The secondary network periodically scans the spectrum once every $T$ seconds and opportunistically transmits over the same frequency band as the primary. 
It is assumed that $T$ exceeds the maximum contention window of the primary network. 
The periodic scanning is in-line with the requirements from present-day CR standards such as the IEEE $802.22$, which mandate that the CRs must periodically scan to detect new primary user activity and freeze their transmissions within a stipulated time period, e.g., within $2$~seconds. 
Hence, the secondary nodes sense the spectrum for a duration $t$ seconds once every $T > t$ seconds. 
If the channel is busy, then the secondary network defers its transmission till the result of the next scan is available. 
If the scan result is idle, the secondary network contends for the channel (along with the primary network) for a period $T-t$ provided it has data to send, after which it senses the spectrum again. 
This is illustrated in Fig.~\ref{fig:timeline}. 
The spectrum sensing by the secondary network ensures that  the secondary nodes contend for the channel only when the primary traffic is low. 
As $t$ is increased, it becomes more and more likely for the primary to start transmission during the scan period, and the probability with which the secondary finds the network idle correspondingly decreases, thereby minimizing the impact on the primary users.
In particular, if $t$ is chosen to exceed the maximum contention window  of the primary network, the secondary network will be able to access the medium only if the primary network has no packets to send. 
In the sequel, we use sensing and scanning interchangeably to refer to the periodic spectrum sensing by the secondary nodes.
We ignore the effect of sensing errors by the secondary network for simplicity and due to lack of space; however, the framework we provide can be extended to handle sensing errors. 
Here, we assume that if there is any overlap between the transmission of the primary nodes and the scan duration $t$ of the secondary nodes, all the secondary nodes sense the spectrum as being busy. 
The synchronism in the secondary network is the same as in any WLAN system, which assumes slot-level synchronization across the nodes. This enables all the nodes in the network to stop and resume their back-off timers nearly simultaneously upon sensing the channel as busy and idle, respectively. 
A detailed discussion on how this can be achieved is presented in \cite{adamis_throughput_2009}. 
Thus, other than the intermittent scanning and deferring of the transmission when the channel is occupied, the secondary network also employs the physical and MAC layer protocols of an IEEE 802.11 based WLAN.  

A simple and accurate Markov model-based analytical formulation of the IEEE 802.11 DCF for the MAC was presented by Bianchi in  \cite{bianchi_performance_2000}. 
In that paper, the back-off state of a given node is modeled to evolve as a slotted discrete-time Markov chain. 
Before the start of each slot, the back-off state transitions to a different state with a certain transition probability. 
These slots are termed transmission slots, owing to the potential transmissions in that slot. 
We refer the reader to \cite{bianchi_performance_2000} for the details of the model, its assumptions and notation. 
A key assumption in that model is that the collision probability remains constant regardless of the back-off stage the nodes are in, due to which, the probability that a node transmits in a given transmission slot (TS) is constant and independent across nodes. 
Thus, in each TS, zero, one or multiple nodes may simultaneously attempt transmission, resulting in an idle slot, a successful transmission slot, or a collision slot, respectively. 
The actual (physical) time duration of the TS thus depends on how many nodes attempt transmission. 
The idle slot is the smallest of all slots, and we refer to its physical time duration as \emph{one real-time slot}. 
The other transmission slots such as a successful transmission slot or a collision slot span multiple real-time slots. 
The fact that the TSs occupy a different number of real-time slots is significant, as it makes analyzing the throughput of the network challenging. 

The collision probability $p$ is defined as the probability with which a transmitted packet at a given node suffers a collision, and it is related to $\tau$, the probability that a particular node transmits in a given TS, by the simultaneous equations \cite{bianchi_performance_2000}:
\begin{align}
  \tau &= \frac{ 2 (1-2p) }{ (1-2p) (1+W) + pW(1-(2p)^{m})}, \label{eq:bianchi_tau} \\
  p &= 1 - (1-\tau)^{N-1}, \label{eq:bianchi1}
\end{align}
where $W$ is the initial contention window length, $m$ is the number of back-off stages, and $N$ is the number of nodes. 
If $\tau$ represents the probability that a given node transmits in a slot, the probability that the remaining $N-1$ nodes remain idle during that slot is simply $\left( 1-\tau \right)^{N-1}$. 
Hence, the packet suffers a collision with probability $1 - \left(1- \tau \right)^{N-1}$, which leads to the relation in (\ref{eq:bianchi1}). 
The relation in (\ref{eq:bianchi_tau}) can be obtained by calculating the steady-state probabilities of the back-off states that result in a transmission.  
The transmission probability determines other important parameters of the network such as its throughput (details are provided in Sec. \ref{pri_sec_thruput}). 
Thus, the challenge in deriving the throughput performance of the primary and secondary networks is in analyzing their transmission probabilities under the coexistence model described above. 
The primary and secondary networks can have different initial contention window lengths and/or different number of back-off stages; we represent them by $W_p$, $W_s$ and $m_p$, $m_s$, respectively. 
Also, we use subscripts $p$ and $s$ to denote primary and secondary network parameters, respectively. 

\begin{remark}
  \label{remark:large_T}
  An assumption in the analysis to follow is that the network is in steady state prior to the scan, due to which, the transmission and collision probabilities are related via (\ref{eq:bianchi_tau}) and (\ref{eq:bianchi1}). 
  This requires $T-t$ to be large. 
  Through simulations, we have observed that if $T-t$ is of the order of $20$ packet durations, then the results provided by our analysis do remain accurate. We note that choosing a large $T-t$ does not necessarily result in a large impact on the primary throughput.
  As mentioned earlier, the primary throughput can be protected to any desired level, by appropriately increasing the value of~$t$. 
  Extending the analysis for a small $T-t$ or for analyzing other link layer metrics such as delay (see \cite{Khabazian-2012}) is out of scope of this work; our focus here is in analyzing the primary and secondary throughput performance.
\end{remark}

The next section presents the analysis of the transmission probabilities when both the primary and secondary networks are saturated, i.e., when both networks always have packets to transmit. 
We extend the analysis to the unsaturated network case in Sec. \ref{sec:unsaturated}.

\section{The Saturated Case} \label{sec:saturated}

When the primary network is saturated, the secondary network will only find the channel idle when all the primary nodes are backing off for the entire scan duration $t$.
At any point in time, the network will be in one of the following states: 1) Only the primary nodes contend for the channel, which happens when the secondary network is sensing for the primary transmission or when the result of the previous scan was busy, and 2) Both primary and secondary nodes  contend for the channel, which happens when the result of the previous scan was idle.
We refer to these states as State-1 and State-2, respectively.
When the network is in State-2, all the nodes in the network contend till the start of the next scan.
We now derive expressions for the transmission and collision probabilities for the network in both states. 

\subsection{State-1}
In this state, only primary nodes are active in the network. 
The transmission probability can therefore be directly obtained from \cite{bianchi_performance_2000}, and the per-node transmission probability ($\tau_{p,1}$) and the collision probability ($p_{p,1}$) are given by \eqref{eq:bianchi_tau} and \eqref{eq:bianchi1} with $\tau_{p,1}$, $p_{p,1}$ and $W_p$ replacing $\tau, p$, and $W$, respectively. 
Note that the subscript $p$ is for the primary, and the subscript $1$ is for State-1. 

\subsection{State-2}
In State-2, all the nodes in the network contend for the channel.
Assuming that the nodes access the channel independently, we can use the Bianchi relationship between the parameters as follows.
Given the collision probabilities $p_{p,2}$ and $p_{s,2}$, the transmission probabilities can be obtained similar to the above, and are given by
\begin{align}
  \tau_{p,2}&=\frac{ 2 (1-2p_{p,2}) }{ (1-2p_{p,2}) (1+W_p) + p_{p,2}W_p(1-(2p_{p,2})^{m_p})}  \label{eq:state2_taup2} \\
  \tau_{s,2}&=\frac{ 2 (1-2p_{s,2}) }{ (1-2p_{s,2}) (1+W_s) + p_{s,2}W_s(1-(2p_{s,2})^{m_s})} . \label{eq:state2_taus2}
\end{align}
Now, $p_{p,2}$ denotes the probability that a transmitted primary packet suffers a collision.
A collision occurs when at least one of the remaining $N_p-1$ primary or one of the $N_s$ secondary nodes also transmit, and is given by
\begin{equation}
  \label{eq:state2_pp2}
  p_{p,2}=1-(1-\tau_{p,2})^{N_p-1}(1-\tau_{s,2})^{N_s},
\end{equation}
and similarly, $p_{s,2}$ is given by
\begin{equation}
  \label{eq:state2_ps2}
  p_{s,2}=1-(1-\tau_{p,2})^{N_p}(1-\tau_{s,2})^{N_s-1}.
\end{equation}
Note that, the interdependence between the primary and secondary network parameters is captured by the relations in \eqref{eq:state2_pp2} and \eqref{eq:state2_ps2}.
To compute $\tau_{p,2}$, $\tau_{s,2}$, $p_{p,2}$ and $p_{s,2}$ it is required to numerically solve the simultaneous equations (\ref{eq:state2_taup2}), (\ref{eq:state2_taus2}), (\ref{eq:state2_pp2}) and (\ref{eq:state2_ps2}). 

The transmission probabilities $\tau_{p,1}, \tau_{p,2}, \tau_{s,2}$ are sufficient to derive the throughput in the two states.
However, the overall throughput depends on the probability  with which the network is in State-1 (or in State-2).
We denote the probability with which the network is in State-1 after the $n$-th scan by $\alpha_{c}[n]$.
$\alpha_{c}[n]$ depends on the scanning duration of the secondary network and has a direct effect on the overall throughput. 

\subsection{The Evaluation of $\alpha_{c}$}
Define the quantities $\alpha_{b}$ and $\alpha_{i}$ as the probabilities with which the current scan result is \emph{busy} conditioned on the event that the previous scan result was \emph{busy} and \emph{idle}, respectively.
The quantities $\alpha_{b}$ and $\alpha_{i}$ depend on the transmission probabilities of the secondary and primary network prior to the scan.
However, by the assumption in Remark \ref{remark:large_T}, the network is in steady-state prior to the scan in either states, and, consequently, the transmission probabilities are as given in the previous sub-section.
Since the transmission probabilities in the two states remain constant for the operation of the network, so do the quantities $\alpha_{b}$ and $\alpha_{i}$.
Below, we provide explicit expressions for $\alpha_{b}$ and $\alpha_{i}$, in terms of the transmission probabilities.

The result of the current scan depends upon the transmission probability of the primary network, which in turn depends on the number of nodes in the network through the collision probability.
The result of the previous scan determines whether the secondary nodes will be active till the current scan, and hence it determines the number of active nodes in the network just prior to the current scan.
We assume that the transmission probability of the primary network remains unchanged for the duration of the scan.
This is because the scan duration is much smaller than the packet duration, and it typically takes several TSs for the network to stabilize to the change from State-2 to State-1. 
Thus, although the secondary nodes defer their transmission, their effect on the primary transmission probability persists during the spectrum scanning phase. 

The probability $\alpha_{c}[n]$ depends on the probability $\alpha_{c}[n-1]$ according to the recursive relation:
\begin{equation}
  \alpha_{c}[n]=\alpha_{b}\alpha_{c}[n-1]+\alpha_{i}(1-\alpha_{c}[n-1]).
\end{equation}
That the sequence $\alpha_c[n]$ converges can be seen from
\begin{equation}
  \alpha_{c}[n]-\frac{\alpha_{i}}{1+\alpha_{i}-\alpha_{b}}=(\alpha_{b}-\alpha_{i})\left[\alpha_{c}[n-1] - \frac{\alpha_{i}}{1+\alpha_{i}-\alpha_{b}} \right] 
\end{equation}
and $\abs{\alpha_{b}-\alpha_{i}} < 1$.
Thus,  $\alpha_{c}[n]$ converges to
\begin{equation}
  \alpha_{c}=\frac{\alpha_{i}}{1+\alpha_{i}-\alpha_{b}}
\end{equation}
The evaluation of $\alpha_{b}$ and $\alpha_{i}$ are thus necessary for the evaluation of $\alpha_{c}$.

\subsubsection{The Evaluation of $\alpha_{b}$}
\label{alpha_b}
Here, prior to the current scan, only primary nodes are active (State-1). 
Hence, the following types of TSs can occur prior to and during the scan:
\begin{itemize}

\item \textbf{Idle slot}: All the primary nodes are in a back-off state.
  Its probability is $p_{i}=(1-\tau_{p,1})^{N_p}$.

\item \textbf{Successful transmission slot}: Exactly one primary node transmits and the other nodes back-off; but any one of the $N_{p}$ nodes can transmit. 
  The probability of this type of slot is $p_{s}=N_{p}\tau_{p,1}(1-\tau_{p,1})^{N_p-1}$.

\item \textbf{Collision slot}: Two or more primary nodes transmit. 
  Its probability is $p_{c}=1-(1-\tau_{p,1})^{N_p}-N_{p}\tau_{p,1}(1-\tau_{p,1})^{N_p-1}$.

\end{itemize}

The different TSs above are also listed in Table~\ref{tab:states_probabilities}. 
Now, we provide some details about the IEEE 802.11 packet transmission that help in enumerating the events that lead to a non-zero overlap between a primary packet transmission and the scan duration, which is necessary in order to evaluate $\alpha_b$.
\begin{table*}[t]
  \begin{center}
    \begin{tabular}{|c|c|c|c|}
      \hline
      State & Description & Probability & Duration\\ \hline
      \multirow{3}{*}{State 1} & Idle & $p_i = \left(1 - \tau_{p,1} \right)^{N_p}$ & 1 \\
      & Successful Primary$^{\dagger}$ & $p_s = N_p \tau_{p,1} \left(1 - \tau_{p,1} \right)^{N_p-1} $ & $\text{TpSuc}+\text{DIFS}$ \\
      & Primary-Primary$^{\ddagger}$   & $p_c = 1 - \left(1 - \tau_{p,1} \right)^{N_p} - N_p \tau_{p,1} \left(1 - \tau_{p,1} \right)^{N_p-1} $ & $\text{TpCol}+\text{EIFS}$ \\ \hline
      \multirow{6}{*}{State 2} & Idle & $q_{ii} = (1-\tau_{p,2})^{N_p}(1-\tau_{s,2})^{N_s}$ & 1\\
      & Successful Primary$^{\dagger}$ & $q_{si} = N_p \tau_{p,2} \left(1 - \tau_{p,2} \right)^{N_p-1} \left( 1 - \tau_{s,2}\right)^{N_s}$ & $\text{TpSuc}+\text{DIFS}$\\
      & Successful Secondary$^{\dagger}$   & $q_{is} = \left( 1 - \tau_{p,2}\right)^{N_p} N_s \tau_{s,2} \left(1 - \tau_{s,2} \right)^{N_s-1} $ & $\text{TsSuc}+\text{DIFS}$ \\
      & Primary-Primary$^{\ddagger}$ & $q_{ci} = \left( 1 - \left(1 - \tau_{p,2} \right)^{N_p} - N_p \tau_{p,2} \left(1 - \tau_{p,2} \right)^{N_p-1} \right) \left(1 - \tau_{s,2} \right)^{N_s}  $ & $\text{TpCol}+\text{EIFS}$\\
      & Secondary-Secondary$^{\ddagger}$ & $ q_{ic} = \left(1 - \tau_{p,2} \right)^{N_p} \left( 1 - \left(1 - \tau_{s,2} \right)^{N_s} - N_s \tau_{s,2} \left(1 - \tau_{s,2} \right)^{N_s-1} \right) $ & $\text{TsCol}+\text{EIFS}$\\
      & Primary-Secondary$^{\ddagger}$ & $q_{cc}=\left( 1 - \left( 1 - \tau_{p,2} \right)^{N_p} \right)\left( 1 - \left( 1 - \tau_{s,2} \right)^{N_s} \right)$ & $\max(\text{TpCol},\text{TsCol})+\text{EIFS}$ \\
      \hline                         
    \end{tabular}
  \end{center}
  \caption{Different transmission slots, their probabilities and real-time slot durations.
    Here, $\dagger$ denotes that the corresponding slot is a successful transmission slot,  $\ddagger$ denotes that it is a collision slot, and the other slots are idle slots.}
  \label{tab:states_probabilities}
\end{table*}

In an IEEE 802.11 network, after each successfully transmitted packet, all the nodes in the network wait for a duration of $\text{DIFS}$ before contending for the channel again.
So, the effective duration of the TS in a successful packet transmission is $\text{TpSuc}+\text{DIFS}$, where $\text{TpSuc}$ is the duration in time for the actual data transmission.
Similarly, after each collision, all the nodes wait for a duration of $\text{EIFS}$ before contending for the channel.
So, the effective duration of the TS in a collision is $\text{TpCol}+\text{EIFS}$. 
$\text{TpCol}$ is the duration in time of the longest packet in collision. 
We normalize all the time durations ($\text{TpSuc}$, $\text{TpCol}$, $\text{EIFS}$, $\text{DIFS}$, $t$, $T$) with respect to the duration of the idle slot, which is the smallest duration slot in the network.

To summarize, there are three kinds of TSs  in the network, with a duration of $1$, $\text{TpSuc}+\text{DIFS}$, $\text{TpCol}+\text{EIFS}$ real-time slots, which occur in a random sequence with probabilities  $p_{i}$, $p_{s}$ and $p_{c}$, respectively. 
The probability that a given real time slot is the start of a TS is 
\begin{equation}
  p_{\text{slot}}=\left[ {p_{s}(\text{TpSuc}+\text{DIFS}) + p_{c}(\text{TpCol}+\text{EIFS}) + p_{i}}\right]^{-1}, 
\end{equation}
since an average duration of a TS is $p_{s}(\text{TpSuc}+\text{DIFS}) + p_{c}(\text{TpCol}+\text{EIFS}) + p_{i}$. 

The result of the spectrum scan of the secondary network will be \emph{idle} in the following mutually exclusive cases. 
In writing the probabilities below, we assume $t>\max(\text{DIFS}, \text{EIFS})$ (we provide the general expression at the end):

a) All the $t$ scan slots are both TS and idle slots. 
The first scan slot will be a TS with probability $p_{\text{slot}}$ and it will be an idle slot given that it is a TS with probability $p_{i}$. 
The successive slots will also be idle transmission slots with probability $p_{i}$. 
Thus, the probability of this event is $p_{\text{slot}}p_{i}^{t}.$

b) The first few of the scan slots are a part of the $\text{DIFS}$ of a previous successful transmission slot and the rest of the scan slots are both TS and idle slots. 
This event can occur in $\text{DIFS}$ number of ways, when one of the primary nodes starts a transmission $\text{TpSuc}+\text{DIFS}-j$ slots prior to the first scan slot; $j = 1, 2, \ldots, \text{DIFS}$. 
Using a similar argument as the above, the probability of this event is 
\begin{equation}
  \sum\limits_{j=1}^{\text{DIFS}} p_{\text{slot}}p_{s}p_{i}^{t-j} = \frac{p_{\text{slot}}p_{s}(p_{i}^{t-\text{DIFS}}-p_{i}^{t})}{1-p_{i}}.
\end{equation}
c) The first few of the scan slots are a part of the $\text{EIFS}$ of a previous collision slot and the rest of the scan slots are both TS and idle slots. Similarly, the probability of this event is given by 
\begin{equation}
  \sum\limits_{j=1}^{\text{EIFS}} p_{\text{slot}}p_{c}p_{i}^{t-j} = \frac{p_{\text{slot}}p_{c}(p_{i}^{t-\text{EIFS}}-p_{i}^{t})}{1-p_{i}}.
\end{equation}
The total probability of all the above events is
\begin{equation}
  1 - \alpha_{b}=p_{\text{slot}}\frac{p_{s}p_{i}^{t-\text{DIFS}}+p_{c}p_{i}^{t-\text{EIFS}}}{p_{s}+p_{c}}.
\end{equation}
If $t < \max(\text{DIFS}, \text{EIFS})$, $\alpha_{b}$ is given by
\begin{equation} 
  \alpha_{b}=1-p_{\text{slot}}\left[\frac{p_{s}p_{i}^{[t_{D}]^{+}}+p_{c}p_{i}^{[t_{E}]^{+}}}{p_{s}+p_{c}} + 
    p_{s}[-t_{D}]^{+} + p_{c}[-t_{E}]^{+} \right]
\end{equation}
where $t_{D}\triangleq t-\text{DIFS}$, $t_{E}\triangleq t-\text{EIFS}$, and $[x]^{+}\triangleq(x+|x|)/{2}$.

\subsubsection{The Evaluation of $\alpha_{i}$}

Here, all the primary and secondary nodes are active (State-2) prior to the current scan. 
Following the development in the previous section, the different TSs that are possible in State-2 of the network are (see also Table~\ref{tab:states_probabilities}): 

\begin{itemize}
\item \textbf{Idle slot}: All the nodes (primary and secondary) back-off.
  Its duration is 1 real-time slot, and it occurs with probability 
  $q_{ii} = (1-\tau_{p,2})^{N_p}(1-\tau_{s,2})^{N_s}$.

\item \textbf{Successful primary transmission slot}: Exactly one primary node transmits, and the rest of the nodes back-off.
  The time duration of this slot is  $\text{TpSuc}+\text{DIFS}$ real-time slots, and its probability is $q_{si} = N_p\tau_{p,2}(1-\tau_{p,2})^{N_p-1}(1-\tau_{s,2})^{N_s}$.

\item \textbf{Successful secondary transmission slot}: Exactly one secondary node transmits and the rest of the nodes back-off. 
  The time duration of this slot is $\text{TsSuc}+\text{DIFS}$ real-time slots, and its probability is $q_{is} = N_s\tau_{s,2}(1-\tau_{p,2})^{N_p}(1-\tau_{s,2})^{N_s-1}$.

\item \textbf{Primary-primary collision slot}: All the secondary nodes are in the back-off state and at least two of the primary nodes transmit. 
  This slot is of duration $\text{TpCol}+\text{EIFS}$ real-time slots, and its probability is  
  \begin{multline}
    q_{ci} = \Big{[} 1 - N_p\tau_{p,2}(1-\tau_{p,2})^{N_p-1} \\
    - (1-\tau_{p,2})^{N_p} \Big{]}  (1-\tau_{s,2})^{N_s}.
  \end{multline}

\item \textbf{Secondary-secondary collision slot}: All the primary nodes are in the back-off state and at least two of the secondary nodes transmit a packet. 
  This slot is of duration $\text{TsCol}+\text{EIFS}$, and its probability is 
  \begin{multline}
    q_{ic} = (1-\tau_{p,2})^{N_p}\Big{[} 1 - N_s\tau_{s,2}(1-\tau_{s,2})^{N_s-1} \\
    - (1-\tau_{s,2})^{N_s} \Big{]}.
  \end{multline} 

\item \textbf{Primary-secondary collision slot}: At least one each of the primary and secondary nodes transmit. 
  The duration and probability of this slot are $\mathrm{max}(\text{TpCol},\text{TsCol})+\text{EIFS}$ and 
  \begin{equation}
    q_{cc} = \left(1-(1-\tau_{p,2})^{N_p}\right)\left(1-(1-\tau_{s,2})^{N_s}\right).
  \end{equation} 
\end{itemize}

Putting the above together, the probability that a given real-time slot is the start of a TS is
\begin{multline}
  q_{\text{slot}} = \Big{[} q_{si}(\text{TpSuc}+\text{DIFS})+q_{is}(\text{TsSuc}+\text{DIFS}) \\
  + q_{ci}(\text{TpCol}+\text{EIFS}) +  q_{ic}(\text{TsCol}+\text{EIFS})  \\
  + q_{cc}(\max(\text{TpCol},\text{TsCol})+\text{EIFS})  + q_{ii} \Big{]}^{-1}.
\end{multline}

As before, the evaluation of $\alpha_{i}$ involves summing the probabilities of the mutually exclusive events in which there is no overlap between any primary transmission and the scan duration. 
However, there are a few important differences in the possible events and in the calculation of their probabilities:
\begin{itemize}

\item If a secondary node wins the contention, but there is insufficient time to transmit the packet before the scan, the node can either defer the packet transmission or transmit a partial packet. 
  We assume that secondary nodes fragment the packets and transmit whatever is possible in the remaining usable time and calculate $\alpha_{i}$ accordingly. 

\item If a scan slot is a TS, the probability of it being an idle slot is not $q_{ii}$, as the secondary nodes do not contend for the channel when scanning. 
  It is sufficient for primary nodes to be in the back-off state for the TS to be idle. 
  The probability of this event is given by
  $q_{i}=(1-\tau_{p,2})^{N_p}$.
\end{itemize}
The expression for the $\alpha_{i}$ is given by:
\begin{multline}
  \alpha_{i} = 1 - q_{\text{slot}}\left\{ q_{i}^{t} + \left[ \frac{q_{i}^{[t_{D}]^{+}} - q_{i}^t}{1-q_{i}} + [-t_{D}]^{+}\right] \left(q_{si}+q_{is}\right) \right.\\
  + (\text{TsSuc}-1)q_{is}q_{i}^{[t_D]^{+}} 
  +   (\text{TsCol}-1)q_{ic}q_{i}^{[t_E]^{+}} \\
  \left. + \left[ \frac{q_{i}^{[t_{E}]^{+}} - q_{i}^t}{1-q_{i}} + [-t_{E}]^{+} \right] \left(q_{ci}+q_{ic}+q_{cc}\right) \right\},
\end{multline} 
where $t_D$, $t_{E}$ and $[x]^{+}$ are as defined before.

\section{The Unsaturated Case} \label{sec:unsaturated}
In this section, we analyze performance of the primary and secondary networks in the unsaturated case, where the nodes do not always have a packet to transmit. 

\subsection{Markov Model and Analysis of the Unsaturated Primary-Only Network}
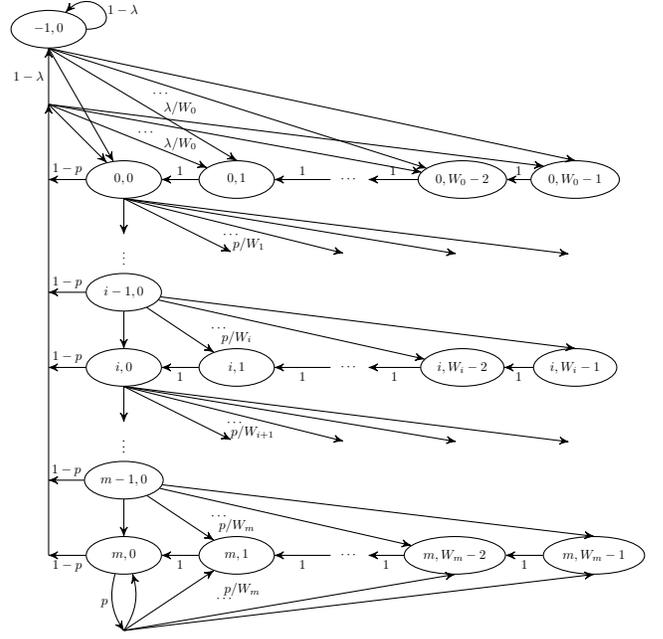
\begin{figure}[t]
  \centering
  \begin{tikzpicture}
[scale=0.5, transform shape, node distance = 12mm, draw=black, thick, >=stealth',
  state/.style={ellipse, thin, draw=black, minimum width=2cm, minimum height=1cm},
  hiddenstate/.style={ellipse, thin, minimum width=0cm, minimum height=0cm}]
  
  \node[state] (s_0_0) at (0,0) {{$0,0$}};
  \node[state] (s_0_1) at (3,0) {{$0,1$}};
  \node[hiddenstate] (s_0_2) at (6,0) {$\cdots$};
  \node[state] (s_0_3) at (9,0) {{$0,W_{0}-2$}};
  \node[state] (s_0_4) at (12,0) {{$0,W_{0}-1$}};

  \draw[->,thin] (s_0_4) -- node[above]{{$1$}} (s_0_3);
  \draw[->,thin] (s_0_3) -- node[above]{{$1$}} (s_0_2);
  \draw[->,thin] (s_0_2) -- node[above]{{$1$}} (s_0_1);
  \draw[->,thin] (s_0_1) -- node[above]{{$1$}} (s_0_0);
  \draw[->,thin] (s_0_0) -- node[above]{{$1-p$}} (-2,0);

  \node[hiddenstate] (s_1_0) at (0,-2) {\small{$\vdots$}};
  \node[hiddenstate] (s_1_1) at (3,-2) {};
  \node[hiddenstate] (s_1_2) at (6,-2) {};
  \node[hiddenstate] (s_1_3) at (9,-2) {};
  \node[hiddenstate] (s_1_4) at (12,-2) {};

  \draw[->,thin] (s_0_0.south) to (s_1_0);
  \draw[->,thin] (s_0_0.south) to node[xshift=1.9cm,yshift=-0.5cm]{{$p/{W_1}$}} node[xshift=1.45cm,yshift=-0.25cm]{$\cdots$} (s_1_1);
  \draw[->,thin] (s_0_0.south) to (s_1_2);
  \draw[->,thin] (s_0_0.south) to (s_1_3);
  \draw[->,thin] (s_0_0.south) to (s_1_4);

  \node[state] (s_i1_0) at (0,-3) {{$i-1,0$}};
  \node[hiddenstate] (s_i1_1) at (3,-3) {};

  \draw[->,thin] (s_i1_0) -- node[above]{{$1-p$}} (-2,-3);

  \node[state] (s_i_0) at (0,-5) {{$i,0$}};
  \node[state] (s_i_1) at (3,-5) {{$i,1$}};
  \node[hiddenstate] (s_i_2) at (6,-5) {$\cdots$};
  \node[state] (s_i_3) at (9,-5) {{$i,W_i-2$}};
  \node[state] (s_i_4) at (12,-5) {{$i,W_i-1$}};

  \draw[->,thin] (s_i1_0) to (s_i_0);
  \draw[->,thin] (s_i1_0) to node[xshift=1.5cm,yshift=-0.2cm]{{$p/{W_i}$}} node[xshift=1.05cm,yshift=0.05cm]{$\cdots$} (s_i_1);
  \draw[->,thin] (s_i1_0) to (s_i_3);
  \draw[->,thin] (s_i1_0) to (s_i_4.north);

  \draw[->,thin] (s_i_4) -- node[below]{{$1$}} (s_i_3);
  \draw[->,thin] (s_i_3) -- node[below]{{$1$}} (s_i_2);
  \draw[->,thin] (s_i_2) -- node[below]{{$1$}} (s_i_1);
  \draw[->,thin] (s_i_1) -- node[below]{{$1$}} (s_i_0);
  \draw[->,thin] (s_i_0) -- node[above]{{$1-p$}} (-2,-5);

  \node[hiddenstate] (s_ip1_0) at (0,-7) {{$\vdots$}};
  \node[hiddenstate] (s_ip1_1) at (3,-7) {};
  \node[hiddenstate] (s_ip1_2) at (6,-7) {};
  \node[hiddenstate] (s_ip1_3) at (9,-7) {};
  \node[hiddenstate] (s_ip1_4) at (12,-7) {};

  \draw[->,thin] (s_i_0.south) to (s_ip1_0);
  \draw[->,thin] (s_i_0.south) to node[xshift=2cm,yshift=-0.5cm]{{$p/W_{i+1}$}} node[xshift=1.55cm,yshift=-0.25cm]{$\cdots$} (s_ip1_1);
  \draw[->,thin] (s_i_0.south) to (s_ip1_2);
  \draw[->,thin] (s_i_0.south) to (s_ip1_3);
  \draw[->,thin] (s_i_0.south) to (s_ip1_4);

  \node[state] (s_m1_0) at (0,-8) {{$m-1,0$}};
  \node[hiddenstate] (s_m1_1) at (3,-8) {};

  \draw[->,thin] (s_m1_0) -- node[above]{{$1-p$}} (-2,-8);

  \node[state] (s_m_0) at (0,-10) {{$m,0$}};
  \node[state] (s_m_1) at (3,-10) {{$m,1$}};
  \node[hiddenstate] (s_m_2) at (6,-10) {$\cdots$};
  \node[state] (s_m_3) at (8.8,-10) {{$m,W_m-2$}};
  \node[state] (s_m_4) at (12.5,-10) {{$m,W_m-1$}};

  \draw[->,thin] (s_m_4) -- node[below]{{$1$}} (s_m_3);
  \draw[->,thin] (s_m_3) -- node[below]{{$1$}} (s_m_2);
  \draw[->,thin] (s_m_2) -- node[below]{{$1$}} (s_m_1);
  \draw[->,thin] (s_m_1) -- node[below]{{$1$}} (s_m_0);
  \draw[->,thin] (s_m_0) -- node[below]{{$1-p$}} (-2,-10);

  \draw[->,thin] (s_m1_0) to (s_m_0);
  \draw[->,thin] (s_m1_0) to node[xshift=1.5cm,yshift=-0.2cm]{{$p/{W_m}$}} node[xshift=1.05cm,yshift=0.05cm]{$\cdots$} (s_m_1);
  \draw[->,thin] (s_m1_0) to (s_m_3);
  \draw[->,thin] (s_m1_0) to (s_m_4.north);
  
  \draw[->,thin] (0,-12) to [bend right=25]  (s_m_0);
  \draw[->,thin] (s_m_0) to [bend right=25]  node[left]{{$p$}} (0,-12);
  \draw[->,thin] (0,-12) -- node[xshift=1.5cm,yshift=0.05cm]{$\cdots$} node[xshift=1.95cm,yshift=0.3cm]{{$p/W_m$}}  (s_m_1);
  \draw[->,thin] (0,-12) -- (s_m_3.south);
  \draw[->,thin] (0,-12) -- (s_m_4.south);

  \node[state] (s_n1_4) at (-2,4) {{$-1,0$}};

  \draw[->,thin] (-2,-10) -- (-2,2);
  \draw[->,thin] (-2,2) -- node[left]{{$1-\lambda$}} (s_n1_4);
  \draw[->,thin] (-2,2) to (s_0_0);
  \draw[->,thin] (-2,2) to node[xshift=1.4cm,yshift=-0.2cm]{{$\lambda/W_0$}} node[xshift=0.5cm,yshift=0.1cm]{$\cdots$} (s_0_1);
  \draw[->,thin] (-2,2) to (s_0_3);
  \draw[->,thin] (-2,2) to (s_0_4.north west);

  \draw[->,thin] (s_n1_4.south) to (s_0_0);
  \draw[->,thin] (s_n1_4.south) to node[xshift=1cm,yshift=-0.1cm]{{$\lambda/W_0$}} node[xshift=0.5cm,yshift=0.3cm]{$\cdots$} (s_0_1.north);
  \draw[->,thin] (s_n1_4.south) to (s_0_3);
  \draw[->,thin] (s_n1_4.south) to (s_0_4.north);

  \draw[->,thin, looseness=4] (s_n1_4) to [in=45,out=0] node[right]{{$1-\lambda$}} (s_n1_4);

\end{tikzpicture}

  \caption{Markov model for nodes in an unsaturated network. In the figure, $W_i=2^{i}W_0$.}
  \label{Unsat_Markov_Model}
\end{figure}

In the literature, several Markov models exist for unsaturated networks \cite{malone_modeling802.11_2007}, \cite{ergen_throughput_2005}.
These models  incorporate the state when nodes do not have a packet to send into an additional idle state.  
In this work, we adopt a different approach, represented by the Markov chain model in Fig.~\ref{Unsat_Markov_Model}.
The advantage of our proposed approach is that it is simple, analytically tractable, and reduces to the Bianchi's model in \cite{bianchi_performance_2000} used in the previous section when the network is fully loaded.
In Fig.~\ref{Unsat_Markov_Model}, the additional state $(-1,0)$  represents the node in unsaturation, i.e., when its data queue is empty. After the successful transmission of a packet, the node will have a new packet in its transmission queue and enter contention again with probability $\lambda$.
Otherwise, with probability $1-\lambda$, it enters the state $(-1,0)$. We denote the quantity $\lambda$ as the traffic intensity.
A $\lambda$ of $1$ corresponds to the saturated network.
If the node enters the state $(-1,0)$, at every TS, a packet arrives in its queue with probability $\lambda$.
In the figure, the $p$ denotes the probability with which the transmitted packet suffers a collision. 
Note that, when $\lambda = 1$, the model reverts to the one in \cite{bianchi_performance_2000} for the saturated network. 

We now derive the steady state transmission and collision probabilities of the unsaturated network.
Let $s_{i,j}$ represent the steady state probability of the state $(i,j)$ in the Markov chain.
Then, it is easy to show that the following equality holds:
\begin{equation}
  s_{i,j} = 
  \begin{cases}
    \frac{1-\lambda}{\lambda}s_{0,0} & \text{if $i=-1, j=0$}\\
    \frac{W_{i}-j}{W_{i}}p^{i}s_{0,0} & \text{if $i \neq m$}\\
    \frac{W_{m}-j}{W_{m}}\frac{p^{m}}{1-p}s_{0,0} & \text{if $i=m$}.
  \end{cases}
\end{equation}
In the above, $m$ denotes the number of back-off stages in the network, and $W_i = 2^i W_0$.
Normalizing the steady state distribution, i.e., since $ \sum\limits_{i=0}^{m} \sum\limits_{j=0}^{W_{i}-1}s_{i,j}=1$, we have
\vspace{-0.1cm} 
\begin{multline}
  s_{0,0}={2(1-2p)(1-p)}\Big{[} (1-2p)(W_0+1) \\
  \quad +pW_0(1-(2p)^{m})+2(1-2p)(1-p)\frac{1-\lambda}{\lambda}\Big{]}^{-1}. 
\end{multline}
The steady state transmission probability $\tau$ is the probability with which the node is in one of the states $\{s_{i,0}: 0 \leq i \leq m \}$. 
That is, 
\begin{multline}
  \label{unsat_tau_p}
  \tau = \sum\limits_{i=0}^{m}s_{i,0}={2(1-2p)}\Big{[}(1-2p)(W_0+1)+pW_0\\
  \times (1-(2p)^{m})+2(1-2p)(1-p)\frac{1-\lambda}{\lambda}\Big{]}^{-1}.
\end{multline}
From the definition of $p$,
\begin{equation}
  p=1-(1-\tau)^{N-1}.
\end{equation}
The two equations above determine the steady state probabilities of the system.

\begin{remark}
  In this work, we do not explicitly model the data queues at the nodes. 
  The value of $\lambda$ is simply representative of the traffic intensity; as it approaches $1$, the model defaults to the saturated case~\cite{bianchi_performance_2000}.
  Another limitation of the model is that it considers the same probability of packet arrival for the different types of TSs, even though their physical duration may be different.
  However, the analysis does account for the probability of occurrence of different types of TSs.
  In the literature, more complex models to represent the unsaturation which consider poisson arrivals (e.g.~\cite{DBLP:journals/corr/abs-1010-4475}) or model  states of fixed real-time length (e.g.~\cite{ZakiElha2004}) have been studied.
  Extending the analysis to these models is not straightforward.
  While our model is simple, it does capture the dynamics of packet transmissions when the primary network is lightly loaded. 
\end{remark}

\subsection{Analysis of the Primary and Secondary Unsaturated Networks}
The evaluation of the parameters $\tau_{p,1}$, $\tau_{p,2}$, $\tau_{s,2}$ and $\alpha_c$, $\alpha_{b}$, $\alpha_{i}$ are similar to the saturated case, with the exception that the relation between $\tau_{p,1}$ and $p_{p,1}$ (and similarly, between $\tau_{p,2}$ and $p_{p,2}$, $\tau_{s,2}$ and $p_{p,2}$) given in (\ref{eq:bianchi_tau}) is now replaced by (\ref{unsat_tau_p}).
Thus, we get:
\begin{align}
  \tau_{p,1} =& {2(1-2p_{p,1})}\Big{[}(1-2p_{p,1})(W_{p}+1)+ p_{p,1}W_{p} \nonumber\\
  & \times (1-(2p_{p,1})^{m_p}) +2(1-2p_{p,1})(1\!-\!p_{p,1})\frac{1-\lambda_p}{\lambda_p}\Big{]}^{-1} \nonumber\\
  p_{p,1}=& 1-(1-\tau_{p,1})^{N_p-1} \nonumber\\
  \tau_{p,2} =&{2(1-2p_{p,2})}\Big{[}(1-2p_{p,2})(W_{p}+1)+p_{p,2}W_{p} \nonumber\\
  & \times (1-(2p_{p,2})^{m_p})+2(1-2p_{p,2})(1\!-\!p_{p,2})\frac{1-\lambda_p}{\lambda_p}\Big{]}^{-1} \nonumber \\
  p_{p,2}=&1-(1-\tau_{p,2})^{N_p-1}(1-\tau_{s,2})^{N_s} \nonumber \\
  \tau_{s,2} =&{2(1-2p_{s,2})}\Big{[}(1-2p_{s,2})(W_{s}+1)+p_{s,2}W_{s} \nonumber\\
  & \times (1-(2p_{s,2})^{m_s})+2(1-2p_{s,2})(1\!-\!p_{s,2})\frac{1-\lambda_s}{\lambda_s}\Big{]}^{-1} \nonumber\\
  p_{s,2}=&1-(1-\tau_{p,2})^{N_p}(1-\tau_{s,2})^{N_s-1}.
\end{align}
In the above, $\lambda_p$ and $\lambda_s$ denote the primary and secondary network traffic intensity according to the unsaturated model in the previous sub-section, respectively.

\section{Primary and Secondary Network Throughput} \label{pri_sec_thruput}
The network throughput is defined as the fraction of the time spent in transmitting  the packets successfully, i.e., after discounting for the time spent in idle and collision slots \cite{bianchi_performance_2000}, \cite{malone_modeling802.11_2007}. 
With this definition, the throughput of the primary network, after averaging over the different states of the network, is given by\footnote{In this article, $\textsl{PT}$ and $\textsl{ST}$  denote the primary and secondary throughput, respectively.}
\begin{equation} \label{eq:thruput1}
  \textsl{PT}=\left( \alpha_{c} p_{\text{slot}} p_{s} + (1-\alpha_{c}) q_{\text{slot}} q_{si} \right) \text{TpSuc}.
\end{equation}
The throughput of the secondary network can be evaluated in State-2, and $1-\alpha_{c}$ determines how often the secondary nodes find the channel available.
The secondary throughput is given by
\begin{equation}\label{eq:thruput2}
  \textsl{ST}=(1-\alpha_c) q_{\text{slot}}q_{is}\,\text{TsSuc}.
\end{equation}

Now, we consider a scheme where the secondary nodes do not scan the spectrum, but periodically observe a mandatory \emph{silence period} of duration $t$ seconds, and contend with the primary for $T-t$ seconds.
Then, the throughput of the primary network is given by
\begin{equation} \label{eq:thruput_pri}
  \textsl{PT}=\left( (1-\beta) p_{\text{slot}} p_{s} + \beta q_{\text{slot}} q_{si} \right) \text{TpSuc},
\end{equation}
where $\beta \triangleq \frac{T-t}{T}$ denotes the fraction of time the secondary network contends with the primary network for transmission of packets, and the other variables are as defined earlier.
Similarly, the secondary throughput is given by
\begin{equation}\label{eq:thruput_sec}
  \textsl{ST}=\beta q_{\text{slot}}q_{is}\text{TsSuc}.
\end{equation}

Finally, the performance of the scheme where the secondary network does not observe any quiet period, but protects the primary network by simply increasing its contention window size can be obtained as a special case of \eqref{eq:thruput_pri} and \eqref{eq:thruput_sec}, by setting $\beta=1$. 
Note that, in this case, the expressions in (\ref{eq:thruput_pri}) and (\ref{eq:thruput_sec}) reduce to the primary and secondary throughput expressions in \cite{He_LCN_2003}, respectively.

\section{Simulation Results}
\label{saturated_simulations}
In this section, we validate the theoretical development in the previous sections using experimental results obtained from the network simulator Ns-2 \cite{ns2}.
We also present the main results of an extensive study of the impact of the sensing duration and contention window size on the primary and secondary throughput. 

\subsection{Setup in Ns-2}

In Ns-2, when a node has a packet to transmit, its physical layer sends the packet to a \emph{channel module}. 
The channel module then schedules the transmission to every neighbor of the transmitter with a pre-calculated propagation delay.
However, the packet is only addressed to one of the neighboring receivers, and thus, all other nodes receive and eventually discard the packet. 
Hence, in Ns-2, the channel ``calls'' or activates a node only when there is a packet to be received by it. 
Due to this, it is not directly feasible for a node to ``turn on'' its receiver and scan the channel within the Ns-2 architecture. 

To allow nodes to sense the channel at will, the channel module is modified to maintain a list of all packets that are in transit at every point in time. 
This information includes the source node address, transmit power, start time, and the total time of the transmission. 
This list is dynamic and has to be updated as and when the packets arrive at and leave the channel. 
Efficient data structures are used for this task and the list is made available to all the nodes. 
Whenever a node wishes to sense the channel, it accesses the packet list and estimates the received power and compares it to a threshold, thereby accomplishing the task of spectrum sensing.

A separate class of cognitive nodes are implemented with the above spectrum sensing capability. 
These nodes synchronously and periodically scan the channel for $t$ seconds. 
We do not consider any sensing errors in the simulations. 
That is, if there is an overlap between the scan duration and any of the primary transmissions, all the secondary nodes freeze their transmission until the channel is sensed as idle again. 
Otherwise, the secondary nodes contend with the primary nodes for the duration $T-t$. 
At the start of the simulation, $N_p$ primary nodes and $N_s$ secondary nodes are placed randomly in an area of size $80$m $\times 80$m. 
The simulation is run for about 500,000 packet transmission attempts. 
Whenever a packet is transmitted, its source, destination, length and the state of the network (State-1 or State-2) are logged into an output file. 
For the unsaturated model, the MAC layer is modified to mimic the analytical model in Section~\ref{sec:unsaturated}.

As shown earlier, the transmission probabilities ($\tau_{p,1}$, $\tau_{p,2}$, $\tau_{s,2}$) of a given node in different network states determine the throughput and delay of the network. 
In any discrete event simulator (particularly, in Ns-2), capturing the idle slots is difficult. 
However, from the transmission log, one can evaluate the number of successful transmissions and the number of collisions in State-1 (simultaneous transmissions by more than one node are collisions) in a given time interval. 
The ratio of these numbers can thus be easily evaluated from the transmission log. 
The theoretical ratio of these numbers (cf. Section \ref{alpha_b}) in State-1 is given by
\begin{equation} \label{eq:psbypc}
  \frac{p_s}{p_c}=\frac{N_{p}\tau_{p,1}(1-\tau_{p,1})^{N_p-1}}{1-N_{p}\tau_{p,1}(1-\tau_{p,1})^{N_p-1}-(1-\tau_{p,1})^{N_p}}.
\end{equation}
Given $N_p$, we numerically solve \eqref{eq:psbypc} to find $\tau_{p,1}$. 
Similarly, the ratio between successful primary transmissions and primary collisions in State-2 ($q_{si}/q_{ci}$) can be used to experimentally evaluate $\tau_{p,2}$. 
Also, for $\tau_{s,2}$ one can use $q_{is}/q_{ic}$. 
We also log the result of each scan of the secondary network. 
The experimental value of $\alpha_{c}$ is the fraction of the times this result is busy.

\subsection{Simulation Study}

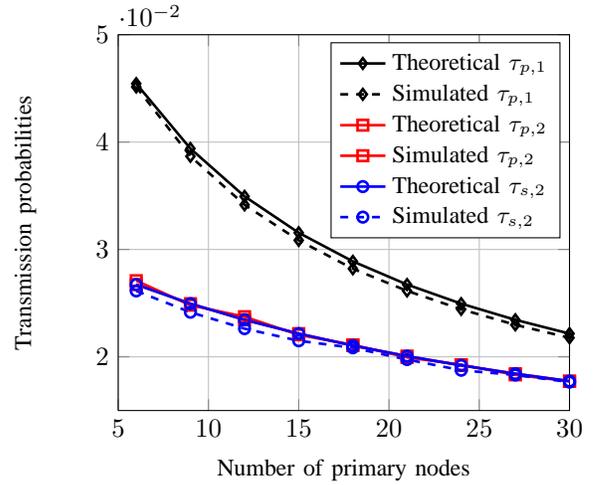
\begin{figure}[t]
\begin{center}
    \setlength\tikzheight{5cm}
    \setlength\tikzwidth{6.0cm}
%
%
\begin{tikzpicture}

\begin{axis}[%
scale only axis,
width=\tikzwidth,
height=\tikzheight,
xmin=5, xmax=30,
ymin=0.015, ymax=0.05,
xlabel={\small{Number of primary nodes}},
ylabel={\small{Transmission probabilities}},
ytick={0.02,0.03,0.04,0.05},
xmajorgrids,
ymajorgrids,
legend entries={\small{Theoretical $\tau_{p,1}$}, \small{Simulated $\tau_{p,1}$}, \small{Theoretical $\tau_{p,2}$}, \small{Simulated $\tau_{p,2}$}, \small{Theoretical $\tau_{s,2}$}, \small{Simulated $\tau_{s,2}$}},
legend style={nodes=right}]

\addplot [
color=black,
solid,
line width=1.0pt,
mark=diamond,
mark options={solid}
]
coordinates{
 (6,0.0454422)(9,0.0393979)(12,0.0349453)(15,0.0315555)(18,0.0288895)(21,0.026734)(24,0.0249512)(27,0.0234487)(30,0.0221627) 
};

\addplot [
color=black,
dashed,
line width=1.0pt,
mark=diamond,
mark options={solid}
]
coordinates{
 (6,0.0451389)(9,0.0386785)(12,0.0341575)(15,0.0308514)(18,0.0282075)(21,0.0261444)(24,0.0244273)(27,0.0229802)(30,0.0217727) 
};

\addplot [
color=red,
solid,
line width=1.0pt,
mark=square,
mark options={solid}
]
coordinates{
 (6,0.026734)(9,0.0249511)(12,0.0234487)(15,0.0221627)(18,0.0210476)(21,0.0200698)(24,0.0192044)(27,0.0184319)(30,0.0177376) 
};

\addplot [
color=red,
solid,
line width=1.0pt,
mark=square,
mark options={solid}
]
coordinates{
 (6,0.0270915)(9,0.0248088)(12,0.0237373)(15,0.0220485)(18,0.0211164)(21,0.0199485)(24,0.0192621)(27,0.0183212)(30,0.0177165) 
};

\addplot [
color=blue,
solid,
line width=1.0pt,
mark=o,
mark options={solid}
]
coordinates{
 (6,0.026734)(9,0.0249512)(12,0.0234487)(15,0.0221627)(18,0.0210476)(21,0.0200698)(24,0.0192043)(27,0.0184319)(30,0.0177375) 
};

\addplot [
color=blue,
dashed,
line width=1.0pt,
mark=o,
mark options={solid}
]
coordinates{
 (6,0.0261398)(9,0.0241547)(12,0.0226287)(15,0.0214966)(18,0.0208428)(21,0.019749)(24,0.0187521)(27,0.0183101)(30,0.017647) 
};

\end{axis}
\end{tikzpicture}

  \caption{Transmission probabilities ($\tau_{p,1}$, $\tau_{p,2}$, $\tau_{s,2}$) as a function of the number of the number of primary nodes in the saturated network case. The number of secondary nodes is $N_s=15$.}
  \label{fig:taup-taus}
  \end{center}
\end{figure}

\begin{figure}[t]
\begin{center}
    \setlength\tikzheight{5cm}
    \setlength\tikzwidth{6.0cm}
%
%
\begin{tikzpicture}

\begin{axis}[%
scale only axis,
width=\tikzwidth,
height=\tikzheight,
xmin=5, xmax=30,
ymin=0.7, ymax=0.95,
xlabel={\small{Number of primary nodes}},
ylabel={\small{$\alpha_{c}$}},
xmajorgrids,
ymajorgrids,
legend entries={\footnotesize{Theoretical, $t=150\mu s$},\footnotesize{Simulated, $t=150\mu s$},\footnotesize{Theoretical, $t=100 \mu s$},\footnotesize{Simulated, $t=100 \mu s$},\footnotesize{Theoretical, $t=50 \mu s$},\footnotesize{Simulated, $t=50 \mu s$}},
legend style={at={(0.98,0.01)},anchor=south east,nodes=right}]

\addplot [
color=black,
solid,
line width=1.0pt,
mark=diamond,
mark options={solid}
]
coordinates{
 (6,0.932894)(9,0.939063)(12,0.938786)(15,0.936939)(18,0.934639)(21,0.932235)(24,0.929856)(27,0.927551)(30,0.925338) 
};

\addplot [
color=black,
dashed,
line width=1.0pt,
mark=diamond,
mark options={solid}
]
coordinates{
 (6,0.931921)(9,0.938094)(12,0.938261)(15,0.942099)(18,0.935425)(21,0.934257)(24,0.926915)(27,0.931754)(30,0.923077) 
};

\addplot [
color=red,
solid,
line width=1.0pt,
mark=square,
mark options={solid}
]
coordinates{
 (6,0.886042)(9,0.904285)(12,0.909692)(15,0.91112)(18,0.910927)(21,0.909969)(24,0.908623)(27,0.907073)(30,0.905419) 
};

\addplot [
color=red,
dashed,
line width=1.0pt,
mark=square,
mark options={solid}
]
coordinates{
 (6,0.86551)(9,0.897047)(12,0.906558)(15,0.913065)(18,0.917737)(21,0.898882)(24,0.911897)(27,0.908393)(30,0.906391) 
};

\addplot [
color=blue,
solid,
line width=1.0pt,
mark=o,
mark options={solid}
]
coordinates{
 (6,0.759618)(9,0.803433)(12,0.822815)(15,0.833134)(18,0.839165)(21,0.842851)(24,0.845127)(27,0.846496)(30,0.847255) 
};

\addplot [
color=blue,
dashed,
line width=1.0pt,
mark=o,
mark options={solid}
]
coordinates{
 (6,0.776406)(9,0.808276)(12,0.834474)(15,0.847489)(18,0.84265)(21,0.843484)(24,0.848657)(27,0.849658)(30,0.858168) 
};

\end{axis}
\end{tikzpicture}

  \caption{Probability of the scan result being busy ($\alpha_c$) in the saturated network case, with $N_s=15$.}
  \label{fig:alpha_c}
  \end{center}
\end{figure}
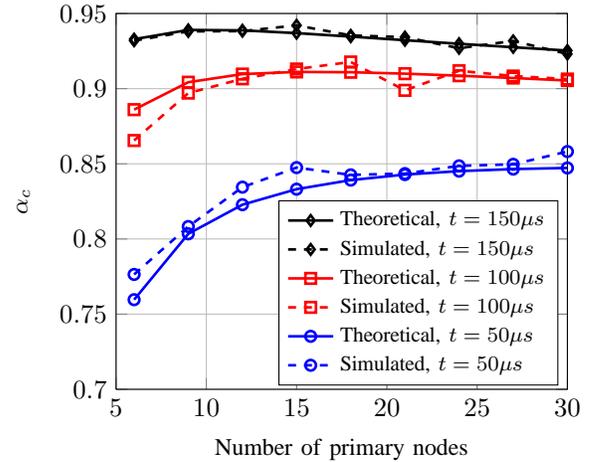

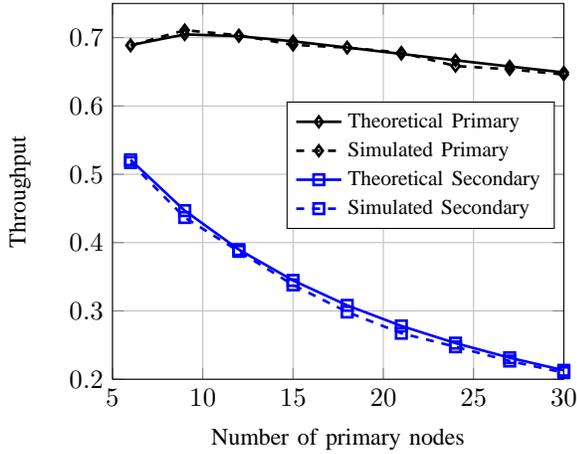
\begin{figure}[t]
\begin{center}
    \setlength\tikzheight{5cm}
    \setlength\tikzwidth{6.0cm}
%
%
\begin{tikzpicture}

\begin{axis}[%
scale only axis,
width=\tikzwidth,
height=\tikzheight,
xmin=5, xmax=30,
ymin=0.2, ymax=0.75,
xlabel={\small{Number of primary nodes}},
ylabel={\small{Throughput}},
xmajorgrids,
ymajorgrids,
legend entries={ \footnotesize{Theoretical Primary}, \footnotesize{Simulated Primary}, \footnotesize{Theoretical Secondary}, \footnotesize{Simulated Secondary} },
legend style={at={(0.98,0.4)},anchor=south east,nodes=right}]

\addplot [
color=black,
solid,
line width=1.0pt,
mark=diamond,
mark options={solid}
]
coordinates{
 (6,0.688982)(9,0.704714)(12,0.702156)(15,0.694595)(18,0.685533)(21,0.676133)(24,0.66683)(27,0.657793)(30,0.649082) 
};

\addplot [
color=black,
dashed,
line width=1.0pt,
mark=diamond,
mark options={solid}
]
coordinates{
 (6,0.688431)(9,0.710905)(12,0.703365)(15,0.689451)(18,0.685332)(21,0.678116)(24,0.658455)(27,0.653689)(30,0.645988) 
};

\addplot [
color=blue,
solid,
line width=1.0pt,
mark=square,
mark options={solid}
]
coordinates{
 (6,0.521148)(9,0.446801)(12,0.389731)(15,0.344629)(18,0.308145)(21,0.278065)(24,0.252868)(27,0.231475)(30,0.213101) 
};

\addplot [
color=blue,
dashed,
line width=1.0pt,
mark=square,
mark options={solid}
]
coordinates{
 (6,0.517241)(9,0.436768)(12,0.387214)(15,0.338117)(18,0.298453)(21,0.267346)(24,0.247565)(27,0.226675)(30,0.209714) 
};

\end{axis}
\end{tikzpicture}

  \caption{Primary and secondary throughput in the saturated network case, with $t=50\mu$s and $N_s=15$.}
  \label{fig:sattput}
\end{center}
\end{figure}

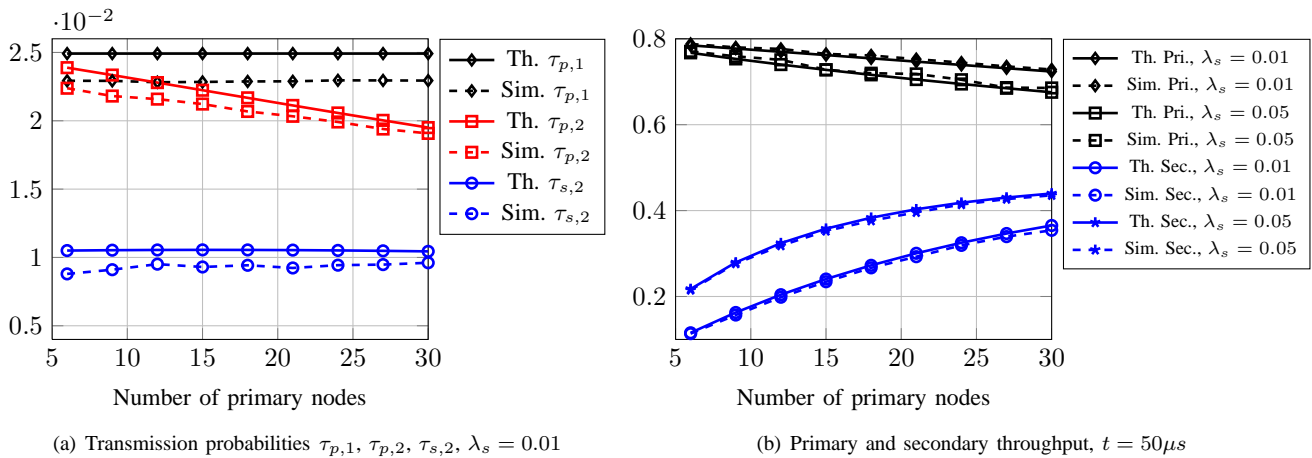
\begin{figure*}[!ht]
  \centering
  \setlength\tikzheight{4.0cm}
  \subfigure[Transmission probabilities $\tau_{p,1}$, $\tau_{p,2}$, $\tau_{s,2}$, $\lambda_{s}=0.01$]{\label{fig:taup_unsat}
%
%
\begin{tikzpicture}

\begin{axis}[%
scale only axis,
width=5cm,
height=\tikzheight,
xmin=5, xmax=30,
ymin=0.004, ymax=0.026,
xlabel={\small{Number of primary nodes}},
xmajorgrids,
ymajorgrids,
legend entries={\small{Th. $\tau_{p,1}$},\small{Sim. $\tau_{p,1}$},\small{Th. $\tau_{p,2}$},\small{Sim. $\tau_{p,2}$},\small{Th. $\tau_{s,2}$},\small{Sim. $\tau_{s,2}$}},
legend pos=outer north east,
]

\addplot [
color=black,
solid,
line width=1.0pt,
mark=diamond,
mark options={solid}
]
coordinates{
 (6,0.0249174)(9,0.0249174)(12,0.0249174)(15,0.0249174)(18,0.0249174)(21,0.0249174)(24,0.0249174)(27,0.0249174)(30,0.0249174) 
};

\addplot [
color=black,
dashed,
line width=1.0pt,
mark=diamond,
mark options={solid}
]
coordinates{
 (6,0.0229271)(9,0.0229215)(12,0.0228208)(15,0.0228426)(18,0.0228785)(21,0.0228887)(24,0.0229606)(27,0.0229529)(30,0.0229417) 
};

\addplot [
color=red,
solid,
line width=1.0pt,
mark=square,
mark options={solid}
]
coordinates{
 (6,0.0238865)(9,0.0233458)(12,0.0227946)(15,0.0222376)(18,0.021679)(21,0.021123)(24,0.0205728)(27,0.0200317)(30,0.0195021) 
};

\addplot [
color=red,
dashed,
line width=1.0pt,
mark=square,
mark options={solid}
]
coordinates{
 (6,0.0223835)(9,0.0218101)(12,0.0215862)(15,0.0212241)(18,0.0206839)(21,0.020329)(24,0.0199139)(27,0.0193984)(30,0.0190776) 
};

\addplot [
color=blue,
solid,
line width=1.0pt,
mark=o,
mark options={solid}
]
coordinates{
 (6,0.0105035)(9,0.0105308)(12,0.0105475)(15,0.0105538)(18,0.01055)(21,0.0105366)(24,0.0105141)(27,0.0104831)(30,0.0104442) 
};

\addplot [
color=blue,
dashed,
line width=1.0pt,
mark=o,
mark options={solid}
]
coordinates{
 (6,0.00879656)(9,0.0091033)(12,0.00950964)(15,0.00930235)(18,0.00942716)(21,0.00923)(24,0.00943472)(27,0.00947664)(30,0.00961198) 
};

\end{axis}
\end{tikzpicture}

}
  \subfigure[Primary and secondary throughput, $t = 50 \mu s$]{\label{fig:tpt_unsat}
%
%
\begin{tikzpicture}

\begin{axis}[%
scale only axis,
width=5cm,
height=\tikzheight,
xmin=5, xmax=30,
ymin=0.1, ymax=0.8,
xlabel={\small{Number of primary nodes}},
xmajorgrids,
ymajorgrids,
legend entries={\scriptsize{Th. Pri., $\lambda_{s}=0.01$}, \scriptsize{Sim. Pri., $\lambda_{s}=0.01$}, \scriptsize{Th. Pri., $\lambda_{s}=0.05$}, \scriptsize{Sim. Pri., $\lambda_{s}=0.05$}, \scriptsize{Th. Sec., $\lambda_{s}=0.01$}, \scriptsize{Sim. Sec., $\lambda_{s}=0.01$}, \scriptsize{Th. Sec., $\lambda_{s}=0.05$}, \scriptsize{Sim. Sec., $\lambda_{s}=0.05$}},
legend pos=outer north east
]

\addplot [
color=black,
solid,
line width=1.0pt,
mark=diamond,
mark options={solid}
]
coordinates{
 (6,0.784699)(9,0.777239)(12,0.769713)(15,0.76213)(18,0.7545)(21,0.746837)(24,0.739152)(27,0.731462)(30,0.72378) 
};

\addplot [
color=black,
dashed,
line width=1.0pt,
mark=diamond,
mark options={solid}
]
coordinates{
 (6,0.78804)(9,0.779988)(12,0.776423)(15,0.764815)(18,0.761193)(21,0.75264)(24,0.744527)(27,0.735679)(30,0.72858) 
};

\addplot [
color=black,
solid,
line width=1.0pt,
mark=square,
mark options={solid}
]
coordinates{
 (6,0.767341)(9,0.753186)(12,0.74002)(15,0.727699)(18,0.716111)(21,0.705163)(24,0.694783)(27,0.684911)(30,0.675496) 
};

\addplot [
color=black,
dashed,
line width=1.0pt,
mark=square,
mark options={solid}
]
coordinates{
 (6,0.77083)(9,0.759104)(12,0.751029)(15,0.728004)(18,0.719839)(21,0.718225)(24,0.704641)(27,0.686401)(30,0.685518) 
};

\addplot [
color=blue,
solid,
line width=1.0pt,
mark=o,
mark options={solid}
]
coordinates{
 (6,0.115486)(9,0.162731)(12,0.204195)(15,0.240601)(18,0.272559)(21,0.300588)(24,0.325142)(27,0.346614)(30,0.365351) 
};

\addplot [
color=blue,
dashed,
line width=1.0pt,
mark=o,
mark options={solid}
]
coordinates{
 (6,0.113492)(9,0.156846)(12,0.198267)(15,0.234381)(18,0.266706)(21,0.292639)(24,0.319018)(27,0.339373)(30,0.354122) 
};

\addplot [
color=blue,
solid,
line width=1.0pt,
mark=star,
mark options={solid}
]
coordinates{
 (6,0.217495)(9,0.279248)(12,0.324243)(15,0.357843)(18,0.383396)(21,0.403094)(24,0.41842)(27,0.430415)(30,0.439824) 
};

\addplot [
color=blue,
dashed,
line width=1.0pt,
mark=star,
mark options={solid}
]
coordinates{
 (6,0.215096)(9,0.276548)(12,0.318339)(15,0.352466)(18,0.376375)(21,0.395967)(24,0.41408)(27,0.426782)(30,0.435631) 
};

\end{axis}
\end{tikzpicture}

}
  \caption{Unsaturated network case. 
    Number of primary nodes $N_p=15$, and $\lambda_{p}=0.05$.}
  \label{fig:alpha}
  \vspace{-0.1in}
\end{figure*}

We now present the network configuration and the simulation results for the \emph{saturated case}. 
The following typical parameters from  IEEE $802.11$ networks were employed in the simulations:  $W_{p}= 32$, $m_{p} =  4$,   $m_{s} = 4$, $\text{TpSuc} = 1178\mu$s, $\text{TsSuc} = 1178\mu$s, $\text{TpCol} =  864\mu$s, $\text{TsCol} = 864\mu$s, $\text{DIFS} = 50\mu$s, $\text{EIFS} = 364\mu$s, $T = 500$ms, Idle Slot Size = 20$\mu$s. 
In Figs. \ref{fig:taup-taus}-\ref{fig:sattput}, we plot the transmission probabilities ($\tau_{p,1}$, $\tau_{p,2}$, $\tau_{s,2}$); the probability of the secondary network finding the spectrum busy ($\alpha_{c}$); and the primary and secondary throughput; respectively. In Fig. \ref{fig:sattput} (and also in Fig. \ref{fig:tpt_unsat}), we omit the $1-\alpha_c$ factor in \eqref{eq:thruput2} and plot the secondary throughput given that it is in State-2, i.e., $q_{\text{slot}} q_{is} \text{TsSuc}$, as we are interested in capturing the variation in secondary throughput with network parameters, rather than in comparing its value to the primary throughput. 
It is clear from the plots that the agreement between the theoretical and Ns-2 simulation-based results is excellent. 
Also, the secondary throughput decreases substantially as the number of primary nodes increases, since a primary transmission during the scan period becomes more and more likely. 
The negative impact on the primary throughput is much less pronounced, which shows the efficacy of the scan period in mitigating the channel capture effect, thereby protecting the throughput of the primary network. 
Thus, the effect of channel capture by the secondary network on the primary network is minimal, provided the secondary network employs DCF with a comparable or higher contention window to access the medium. 

\begin{figure}[t]
  \begin{center}
    \setlength\tikzwidth{6.0cm}
    \begin{tikzpicture}

\begin{axis}[%
scale only axis,
width=\tikzwidth,
height=1.5cm,
xmin=0, xmax=6e-4,
xtick={0e-4,1e-4,2e-4,3e-4,4e-4,5e-4,6e-4},
scaled x ticks={base 10:4},
ylabel={\small{Throughput}},
ymin=0, ymax=0.8,
xmajorgrids,
ymajorgrids,
legend columns=2,
legend entries={ \footnotesize{Theoretical Primary}, \footnotesize{Simulated Primary}, \footnotesize{Theoretical Secondary}, \footnotesize{Simulated Secondary} },
legend to name=commonlegend_TPut]

\addplot [
color=black,
solid,
line width=1.0pt,
mark=square,
mark options={solid}
]
coordinates{
 (2e-05,0.665795)(6e-05,0.689545)(0.0001,0.709353)(0.00014,0.722597)(0.00018,0.728566)(0.0003,0.735316)(0.0004,0.745496)(0.0005,0.749454)(0.0006,0.74955) 
};

\addplot [
color=black,
dashed,
line width=1.0pt,
mark=square,
mark options={solid}
]
coordinates{
 (2e-05,0.672654)(6e-05,0.698774)(0.0001,0.720623)(0.00014,0.729773)(0.00018,0.735099)(0.0003,0.745671)(0.0004,0.753317)(0.0005,0.757754)(0.0006,0.75815) 
};

\addplot [
color=blue,
solid,
line width=1.0pt,
mark=diamond,
mark options={solid}
]
coordinates{
 (2e-05,0.06816)(6e-05,0.0489032)(0.0001,0.0325196)(0.00014,0.0221162)(0.00018,0.0170572)(0.0003,0.0114832)(0.0004,0.00339107)(0.0005,0.000340141)(0.0006,0) 
};

\addplot [
color=blue,
dashed,
line width=1.0pt,
mark=diamond,
mark options={solid}
]
coordinates{
 (2e-05,0.0698387)(6e-05,0.0485111)(0.0001,0.0306709)(0.00014,0.0232003)(0.00018,0.0188513)(0.0003,0.0102191)(0.0004,0.00397549)(0.0005,0.000352978)(0.0006,2.90749e-05) 
};

\end{axis}
\end{tikzpicture}

\begin{tikzpicture}

\begin{axis}[%
scale only axis,
width=\tikzwidth,
height=1.5cm,
xlabel={\small{Scanning duration $(t)$}},
ylabel={\small{Throughput}},
xmin=2e-5, xmax=10e-5,
ymin=0, ymax=0.8,
xtick={2e-5,4e-5,6e-5,8e-5,10e-5},
xmajorgrids,
ymajorgrids]

\addplot [
color=black,
solid,
line width=1.0pt,
mark=square,
mark options={solid}
]
coordinates{
 (2e-05,0.68316)(4e-05,0.692865)(8e-05,0.714861)(0.0001,0.720742) 
};

\addplot [
color=blue,
solid,
line width=1.0pt,
mark=diamond,
mark options={solid}
]
coordinates{
 (2e-05,0.0511453)(4e-05,0.0436795)(8e-05,0.0263467)(0.0001,0.0217519) 
};

\addplot [
color=blue,
dashed,
line width=1.0pt,
mark=diamond,
mark options={solid}
]
coordinates{
 (2e-05,0.0697607)(4e-05,0.0614075)(8e-05,0.0371562)(0.0001,0.0304995) 
};

\addplot [
color=black,
dashed,
line width=1.0pt,
mark=square,
mark options={solid}
]
coordinates{
 (2e-05,0.67275)(4e-05,0.68298)(8e-05,0.712681)(0.0001,0.720833) 
};

\end{axis}
\end{tikzpicture}

\ref{commonlegend_TPut}

  \caption{Primary and secondary throughput as a function of the scan duration with $N_p=16$, $N_s=16$, $W_s=32$; saturated case; 
    $T=500$ms in the top plot and $T=23.5$ms (around $20$ packet durations) in the bottom plot. 
    If $T$ is further decreased, the gap between theoretical and simulated values increases.}
  \label{fig:tput_t}
  \end{center}
\end{figure}
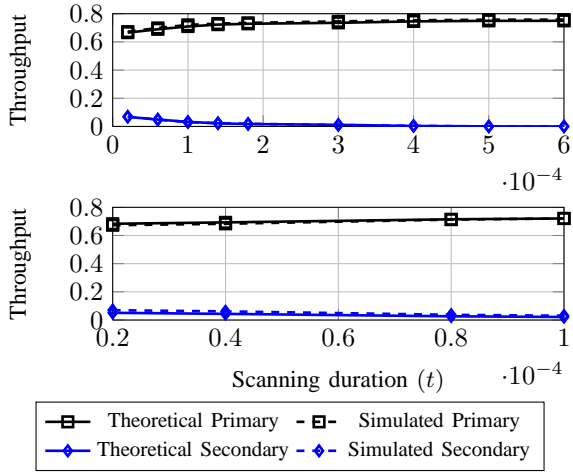

The effect of the scan duration is illustrated in Fig. \ref{fig:tput_t} with $T=500$ms (top plot) and $T=23.5$ms (or $20$ packet durations) (bottom plot).
As the scanning duration is increased, it becomes unlikely for the secondary network to find the spectrum as idle. 
Hence, beyond  $t = 400 \mu$s in Fig. \ref{fig:tput_t} (top plot), the secondary and primary throughput values saturate, and are relatively insensitive to the scanning duration. 
We note that the agreement between theoretical and simulated throughput is good even at $T=23.5$ms, however, the gap between these will increase as $T$ is further reduced. 
We also notice that the secondary throughput is significantly smaller than that of the primary, even for very low scanning durations. This is because the secondary network stays idle for a relatively long time whenever the scan result is busy. 
Moreover, the timing of the scan duration is not synchronized with an end-of-transmission of the primary network, due to which, the probability of the secondary network scanning and capturing the channel between primary transmissions is low. 
Again, this shows that the effect of channel capture on the primary network is minimal, when the secondary network also uses the IEEE 802.11~DCF. 

In the \emph{unsaturated case}, we provide the simulation results of the transmission probabilities in Fig.~\ref{fig:taup_unsat} and the primary and secondary throughput for different unsaturation levels of the secondary network $\lambda_s$ in Fig.~\ref{fig:tpt_unsat}. 
From Fig.~\ref{fig:taup_unsat}, we see that as the number of secondary nodes increases, the transmission probabilities $\tau_{p,2}$ and $\tau_{s,2}$ closely match the theoretical expressions. 
There is a small gap between the theoretical and experimental curves for the transmission probability $\tau_{p,1}$. 
This is because the assumption that the collision probability is a constant independent of the back-off stage, which is necessary for analytical tractability, is not valid when the number of nodes is small and its effect is even more pronounced when the network is highly unsaturated. 
However, the simulation results for the throughput match the theoretical expressions well, and hence the deviation in the $\tau_{p,1}$ curve is not significant.

\subsection{Primary-Secondary Throughput Trade-off}
In this subsection, we study and compare different mechanisms for constraining the maximum impact of the secondary network on the primary network throughput. 
We consider saturated networks, and determine the combination of secondary parameters that result in the maximum secondary throughput, subject to an upper bound on the maximum loss of primary throughput (say, $10\%$) due to the secondary network. 
The optimization requires the secondary to know the values of $N_p$, $\lambda_p$, $W_p$, and $m_p$. 
Usually, initial contention window length $W_p$ is fixed for a given protocol (e.g., in $802.11$ networks, it is $32$). 
The performance is insensitive to $m_p$ as long as it is $\geq 3$. It is also possible to estimate the number of primary nodes and the traffic intensity in an $802.11$ network (for e.g.,~see \cite{BianTinn2003}, \cite{ToleVerc2006}). 
In this subsection, we assume the perfect knowledge of $N_p$, $W_p$ and $m_p$ and optimize the secondary parameters. 
In the next sub-section, we study the effect of misestimation of the primary network parameters on the primary and secondary throughput performance.

For the sensing strategy, increasing $W_s$ causes the secondary network to spend more time in back-off, thereby allowing the primary network to win contentions with higher probability, while increasing $t$ causes the secondary network to find the channel busy with higher probability, and hence stay silent for a longer duration. Hence, the optimization problem can be written as
\begin{align}
  \label{eq:optimization_sensing}
  &  \qquad \max_{t,W_s} \qquad \left( 1 - \alpha_c \right) q_{\text{slot}} q_{is} \text{TsSuc} \notag \\
  &  \text{subject to} \quad \alpha_c p_{\text{slot}} p_s  + \left( 1 - \alpha_c \right) q_{\text{slot}} q_{si} \leq 0.9\,  p_{\text{slot}} p_s.
\end{align}
The primary throughput without secondary network is given by the throughput in State-1, which is $p_{\text{slot}} p_s \text{TpSuc}$. 
The throughput in the presence of the secondary network is given by (\ref{eq:thruput1}), and hence, requiring an at most $10 \%$ loss in the primary throughput leads to the constraint in (\ref{eq:optimization_sensing}).
Similarly, the optimization problem for the mandatory silence period strategy with the primary and secondary throughput given by (\ref{eq:thruput_pri}) and (\ref{eq:thruput_sec}) can be written as
\begin{align}
  \label{eq:optimization_silence}
  &  \qquad \max_{\beta,W_s} \qquad \beta q_{\text{slot}} q_{is} \text{TsSuc} \notag \\
  &  \text{subject to} \quad \left( 1 - \beta \right) p_{\text{slot}} p_s  + \beta q_{\text{slot}} q_{si} \leq 0.9\,  p_{\text{slot}} p_s .
\end{align}
Finally, for the strategy of increasing secondary contention window length, it is obtained by substituting $\beta=1$ above, to get
\begin{align}
  \label{eq:optimization_coexistence}
  &  \qquad \max_{W_s} \qquad  q_{\text{slot}} q_{is} \text{TsSuc} \notag \\
  &  \text{subject to} \quad q_{\text{slot}} q_{si} \leq 0.9\, p_{\text{slot}} p_s .
\end{align}

Although we derived expressions for the primary and secondary throughput, it is difficult to directly solve for the optimal parameters for the problems in (\ref{eq:optimization_sensing}), (\ref{eq:optimization_silence}) and (\ref{eq:optimization_coexistence}) in closed form. 
We therefore find the optimal parameters by performing a \emph{numerical search} over a range of values for $t$, $W_s$ and $\beta$. 
The results of this search are presented in Table~\ref{tab:thp}. 
In this table, $\textsl{PT}$ denotes the primary throughput and $\textsl{ST}$ denotes the secondary throughput. 
When optimized over the entire range of the respective parameters, all the three schemes perform nearly equally well. 
Thus, the optimized coexistence scheme is advantageous in practice, due to its simplicity, when the networks are saturated and a perfect knowledge of primary network parameters is available.
\begin{table}[t]
  \begin{center}
    \begin{tabular}{ | p{1.6cm} | p{1.6cm} | p{1.6cm} | p{1.6cm} | }
      \hline
      & Scanning & Coexisting Primary and Secondary & Mandatory Silent Period \\ \hline
      $N_p=16$,   &  $t=10\mu S$,   &   &  $\beta=0.70$,\\ 
      $N_s=4$,  &  $W_s=11$,  &  $W_s=80$, &   $W_s=54$,\\ 
      $\textsl{PT}=0.682$ & $\textsl{ST}=0.064$ & $\textsl{ST}=0.065$ & $\textsl{ST}=0.065$ \\ \hline
      
      $N_p=16$, & $t=5\mu S$,  &   & $\beta=0.85$,   \\
      $N_s=8$, & $W_s=21$, &   $W_s=158$,&  $W_s=132$,  \\
      $\textsl{PT}=0.682$ & $\textsl{ST}=0.063$  &  $\textsl{ST}=0.065$ &   $\textsl{ST}=0.065 $\\ \hline
      
      $N_p=16$,  &  $t=20\mu S$,  &  &  $\beta=1$, \\ 
       $N_s=16$, &  $W_s=37$, & $W_s=314$, &   $W_s=314$,  \\ 
      $\textsl{PT}=0.682$ & $\textsl{ST}=0.062$  &  $\textsl{ST}=0.065$ & $\textsl{ST}=0.065$   \\ \hline  
      
      $N_p=32$,   &  $t=20\mu S$,  &   &  $\beta=0.90$, \\ 
       $N_s=4$,  &   $W_s=6$,  & $W_s=43$,  &  $W_s=38$,\\ 
      $\textsl{PT}=0.613$ & $\textsl{ST}=0.056$   & $\textsl{ST}=0.056$  & $\textsl{ST}=0.057$\\ \hline
      
      $N_p=32$, &  $t=10\mu S$,  &    &  $\beta=1$, \\
       $N_s=8$,&   $W_s=12$,  &  $W_s=84$,  &  $W_s=84$,\\
      $\textsl{PT}=0.613$ & $\textsl{ST}=0.054$  & $\textsl{ST}=0.057$ & $\textsl{ST}=0.057$\\ \hline

     $N_p=32$, &  $t=10\mu S$,   &    &  $\beta=1$,\\
     $N_s=16$,&   $W_s=23$,  &  $W_s=167$,  &   $W_s=167$,\\
      $\textsl{PT}=0.613$ & $\textsl{ST}=0.054$  & $\textsl{ST}=0.057$ & $\textsl{ST}=0.057$\\ \hline
    \end{tabular}
  \end{center}
  \caption{ Comparison of the optimized performance of the different MAC layer schemes for  secondary access. $\textsl{PT}$ stands for primary throughput and $\textsl{ST}$ stands for secondary throughput. With $N_p=16$, $90\%$ of the maximum primary throughput is $0.682$, and with $N_p=32$, it is~$0.613$.}
  \label{tab:thp}
\end{table}

\subsection{Robustness to Primary Network Parameters}

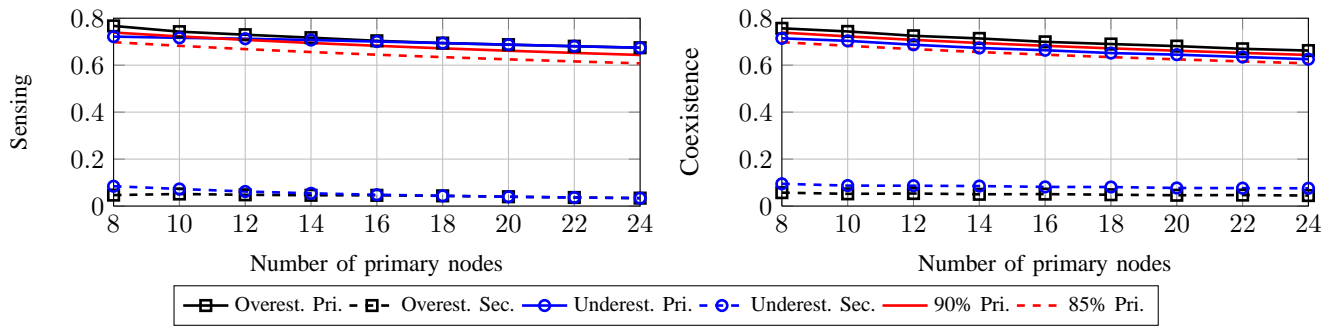
\begin{figure*}[t]
  \begin{center}
    \begin{tikzpicture}

\begin{axis}[%
scale only axis,
width=7cm,
height=2.5cm,
xmin=8, xmax=24,
ymin=0, ymax=0.8,
xlabel={\small{Number of primary nodes}},
ylabel={\small{Sensing}},
xmajorgrids,
ymajorgrids,
legend columns=-1,
legend entries={ \footnotesize{Overest. Pri.}, \footnotesize{Overest. Sec.}, \footnotesize{Underest. Pri.}, \footnotesize{Underest. Sec.}, \footnotesize{90\% Pri.}, \footnotesize{85\% Pri.} },
legend to name=commonlegend_robustness_nodes]

\addplot [
color=black,
solid,
line width=1.0pt,
mark=square,
mark options={solid}
]
coordinates{
 (8,0.767011)(10,0.742607)(12,0.729627)(14,0.716969)(16,0.703644)(18,0.693925)(20,0.687231)(22,0.680604)(24,0.674104) 
};

\addplot [
color=black,
dashed,
line width=1.0pt,
mark=square,
mark options={solid}
]
coordinates{
 (8,0.0467529)(10,0.0514363)(12,0.0480461)(14,0.0460071)(16,0.0458204)(18,0.0435894)(20,0.039751)(22,0.0365564)(24,0.033854)
};

\addplot [
color=blue,
solid,
line width=1.0pt,
mark=o,
mark options={solid}
]
coordinates{
 (8,0.721499)(10,0.716365)(12,0.712506)(14,0.706907)(16,0.700566)(18,0.693925)(20,0.687231)(22,0.680604)(24,0.674104)
};

\addplot [
color=blue,
dashed,
line width=1.0pt,
mark=o,
mark options={solid}
]
coordinates{
 (8,0.0844758)(10,0.0728208)(12,0.0620745)(14,0.0542703)(16,0.0483066)(18,0.0435894)(20,0.039751)(22,0.0365564)(24,0.033854)
};

\addplot [
color=red,
solid,
line width=1.0pt
]
coordinates{
 (8,0.73885)(10,0.722349)(12,0.707586)(14,0.694343)(16,0.682369)(18,0.671439)(20,0.661387)(22,0.652073)(24,0.643394)
};

\addplot [
color=red,
dashed,
line width=1.0pt
]
coordinates{
 (8,0.697803)(10,0.682218)(12,0.668276)(14,0.655768)(16,0.64446)(18,0.634137)(20,0.624643)(22,0.615847)(24,0.60765) 
};

\end{axis}
\end{tikzpicture}
\begin{tikzpicture}

\begin{axis}[%
scale only axis,
width=7cm,
height=2.5cm,
xmin=8, xmax=24,
xlabel={\small{Number of primary nodes}},
ylabel={Coexistence},
ymin=0, ymax=0.8,
xmajorgrids,
ymajorgrids,
]

\addplot [
color=black,
solid,
line width=1.0pt,
mark=square,
mark options={solid}
]
coordinates{
 (8,0.756779)(10,0.743502)(12,0.724981)(14,0.713727)(16,0.699212)(18,0.689922)(20,0.68102)(22,0.669609)(24,0.662099)
};

\addplot [
color=black,
dashed,
line width=1.0pt,
mark=square,
mark options={solid}
]
coordinates{
 (8,0.0571575)(10,0.0521026)(12,0.0535367)(14,0.0502053)(16,0.0509717)(18,0.0482812)(20,0.0461289)(22,0.0468353)(24,0.0448426)
};

\addplot [
color=blue,
solid,
line width=1.0pt,
mark=o,
mark options={solid}
]
coordinates{
 (8,0.714178)(10,0.703451)(12,0.686884)(14,0.673084)(16,0.663471)(18,0.65147)(20,0.644664)(22,0.634691)(24,0.625378)
};

\addplot [
color=blue,
dashed,
line width=1.0pt,
mark=o,
mark options={solid}
]
coordinates{
 (8,0.094832)(10,0.0871664)(12,0.0866113)(14,0.0852629)(16,0.081624)(18,0.081093)(20,0.0770218)(22,0.0763671)(24,0.0757642)
};

\addplot [
color=red,
solid,
line width=1.0pt
]
coordinates{
 (8,0.73885)(10,0.722349)(12,0.707586)(14,0.694343)(16,0.682369)(18,0.671439)(20,0.661387)(22,0.652073)(24,0.643394)
};

\addplot [
color=red,
dashed,
line width=1.0pt
]
coordinates{
 (8,0.697803)(10,0.682218)(12,0.668276)(14,0.655768)(16,0.64446)(18,0.634137)(20,0.624643)(22,0.615847)(24,0.60765) 
};
\end{axis}
\end{tikzpicture}

\ref{commonlegend_robustness_nodes}

    \caption{Primary and secondary throughput under a $30\%$ mismatch between the actual number of primary nodes (the $x$-axis) and the number of nodes assumed by the secondary for optimizing its transmission parameters. }
    \label{fig:robustness_nodes}
\end{center}
\end{figure*}

In this subsection, we study the impact of imperfect knowledge about the primary network on the throughputs of both the primary and secondary networks. 
We consider a $30\%$ mismatch between the number of primary nodes deployed and the number of primary nodes assumed by the secondary network for optimizing its scan duration and back-off length. 
The results are presented in Fig.~\ref{fig:robustness_nodes}. 
The curves corresponding to $90\%$ primary throughput (the design target) and $85\%$ primary throughput are also shown for reference.
We observe that underestimating the number of primary nodes is better for the primary throughput. 
On the other hand, overestimation negatively impacts its throughput. 
This is because overestimation leads to a more aggressive selection of secondary network parameters in the optimization. 
Nonetheless, even with an estimation error of $30\%$, the impact on primary throughput exceeds the designed value by less than $5\%$ with all the schemes. 
When the number of primary nodes is large ($\geq 15$), the sensing strategy is the most robust of the three schemes. 
The performance of the mandatory silence period strategy is very similar to the coexisting strategy and it is therefore not shown. 

We also consider the effect of traffic mismatch at the secondary nodes. 
In particular, we simulate an extreme case, where the secondary network optimizes its parameters by assuming that the primary is saturated, when, in reality, the primary network is unsaturated with a traffic intensity of $\lambda_p = 0.001$. 
Table \ref{tab:thp1} lists the the primary and secondary throughputs achieved. 
We see that the mandatory silent period strategy offers the best protection to primary network under the traffic mismatch. 
Between the sensing and the coexistence strategy with larger contention window, the latter strategy is more robust to the traffic mismatch.
\begin{table}[t]
  \begin{center}
    \begin{tabular}{ | p{1.5cm} | p{1.5cm} | p{1.5cm} | p{1.5cm} | }   \hline
      \vspace{0.1cm}$(N_p, N_s)$& Scanning & Coexisting Primary and Secondary & Mandatory Silent Period \\ \hline
      \vspace{0.01cm}$(16,4)$  &  $\textsl{PT}=0.130$, $\textsl{ST}=0.676$  &  $\textsl{PT}=0.133$, $\textsl{ST}=0.702$ &  $\textsl{PT}=0.220$, $\textsl{ST}=0.525$ \\  \hline
      
     \vspace{0.01cm} $(16, 8)$ & $\textsl{PT}=0.097$, $\textsl{ST}=0.684$ &   $\textsl{PT}=0.133$, $\textsl{ST}=0.698$ & $\textsl{PT}=0.175$, $\textsl{ST}=0.613$  \\ \hline
      
      \vspace{0.01cm}$(16,16)$ &  $\textsl{PT}=0.079$, $\textsl{ST}=0.649$ & $\textsl{PT}=0.133$, $\textsl{ST}=0.696$ &  $\textsl{PT}=0.133$, $\textsl{ST}=0.696$  \\ \hline
    \end{tabular}
  \end{center}
  \caption{Throughput comparison of the different MAC layer schemes in the presence of mismatched traffic intensity, with an unsaturated primary network and with $\lambda_p = 0.001$. 
    The secondary network is saturated and is designed assuming that the primary network is saturated. 
    $\textsl{PT}$ denotes the primary throughput and $\textsl{ST}$ denotes the secondary throughput. 
    In the absence of the secondary, $90\%$ of the maximum primary throughput with $\lambda_p = 0.001$ is~$0.444$.}
  \label{tab:thp1}
\end{table}

\section{Conclusions} \label{sec:conclusions}
In this paper, we analyzed the coexistence of a secondary network, which frequently scans for the availability of the spectrum, with a primary WLAN network. 
We characterized the probability with which the secondary network is able to access the channel and the primary and secondary throughput, as a function of the scan duration. 
We considered both saturated and unsaturated networks. 
In the latter case, we developed a new Markov chain model for analyzing the performance of unsaturated networks, and derived analytical expressions for the primary and secondary throughput. 
We validated our analysis using extensive  simulations on the network simulator, Ns-2. 
We showed that channel captures by the secondary network in fact does not have a significant impact on the primary network, provided the scan duration is at least of the order of a few idle slots. 
For example, periodically scanning the spectrum for just three idle slots ensures that the loss in primary throughput is less than $10\%$, when the primary and secondary networks have the same number of nodes and contention window sizes. 
Moreover, increasing the secondary contention window size is a simple method, and is only marginally inferior to the other complex methods, to maximize the secondary network performance, while protecting the primary throughput.
One scenario where the sensing based scheme could outperform the other schemes is the bursty traffic case. 
Periodic sensing to detect traffic bursts, and using a short contention window between the bursts could potentially offer the best performance. 
Further, one could envision using the periodic sensing to estimate primary traffic parameters, use model-based prediction of spectrum availability, etc., thereby exploiting \emph{all} the information available from the sensing outcomes. 
These ideas offer interesting avenues for future work in designing and optimizing CR networks. 

\section*{Acknowledgements}
The authors thank the anonymous reviewers for their detailed comments and suggestions which significantly improved the presentation of the article.

\bibliographystyle{IEEEtran}
\bibliography{References_ThruputStudy.bib}

\end{document}